\documentclass[journal]{IEEEtran}
\usepackage{amssymb,amsfonts,cite}
\usepackage{amsthm, multirow}
\usepackage{graphicx}
\usepackage[cmex10]{amsmath}
\usepackage{algorithmic, algorithm}
\usepackage{array}
\usepackage{stfloats}
\usepackage[square, comma, numbers, sort&compress]{natbib}
\usepackage{arydshln}
\usepackage{float}
\usepackage{epstopdf}
\usepackage {verbatim}

\theoremstyle{definition, italic}
\newtheorem{prop}{Proposition}

\renewcommand{\tablename}{\textsc{Algorithm}} 

\title{Energy Minimization for the Half-Duplex Relay Channel with Decode-Forward Relaying }

\author{Fanny Parzysz, \IEEEmembership{Student Member, IEEE}, Mai Vu, \IEEEmembership{Member, IEEE}, Fran\c cois Gagnon, \IEEEmembership{Senior Member IEEE}%
\thanks{This work has been supported in part by Ultra Electronics TCS and the Natural Science and Engineering Council of Canada as part of the “High Performance Emergency and Tactical Wireless Communication Chair” at \'{E}cole de Technologie Sup\'{e}rieure. A part of this work was presented at the IEEE Information Theory Workshop 2011.}
\thanks{Fanny Parzysz and Fran\c cois Gagnon are with \'Ecole de Technologie Sup\'erieure, Montreal, Canada. Mai Vu was with McGill University, she is now with the  Electrical and Computer Engineering department at Tufts University. (emails: Fanny.Parzysz@lacime.etsmtl.ca; Mai.Vu@tufts.edu; Francois.Gagnon@etsmtl.ca) }
}

\begin{document}
\maketitle

\begin{abstract}
We analyze coding for energy efficiency in relay channels at a fixed source rate. We first propose a half-duplex decode-forward coding scheme for the Gaussian relay channel. We then derive three optimal sets of power allocation, which respectively minimize the network, the relay and the source energy consumption. These optimal power allocations are given in closed-form, which have so far remained implicit for maximum-rate schemes. Moreover, analysis shows that minimizing the network energy consumption at a given rate is not equivalent to maximizing the rate given energy, since it only covers part of all rates achievable by decode-forward.
We thus combine the optimized schemes for network and relay energy consumptions into a generalized one, which then covers all achievable rates. This generalized scheme is not only energy-optimal for the desired source rate but also rate-optimal for the consumed energy. 
The results also give a detailed understanding of the power consumption regimes and allow a comprehensive description of the optimal message coding and resource allocation for each desired source rate and channel realization.
Finally, we simulate the proposed schemes in a realistic environment, considering path-loss and shadowing as modelled in the 3GPP standard. Significant energy gain can be obtained over both direct and two-hop transmissions, particularly when the source is far from relay and destination.
\end{abstract}

%\begin{IEEEkeywords}
%Relay channel, Half-duplex, Decode-Forward, Energy efficiency, Rate maximization
%\end{IEEEkeywords}
\bstctlcite{IEEEexample:BSTcontrol}
\newcounter{MYtempeqncnt}

\section{Introduction}
\label{introduction}

As a key feature of future wireless systems, relaying has attracted significant attention in recent years and a rich literature has contributed to analyzing the capacity of relay channels. 
Upper and lower bounds were first established in \cite{capacityRelayChannel}, and derived for AWGN relay channels considering energy-per-bit in \cite{art4-5}. More recently, bounds have been extended to multiple relays in \cite{art4-9} and MIMO channels in \cite{wang2005,simoens2008}. Considering the three-node networks, power allocation maximizing the source rate at a given power has been extensively analyzed in \cite{Host-Madsen,Host-MadsenJournal}.

In practice, however, user applications are mainly associated with a fixed minimum rate, defined as a Quality-of-Service feature \cite{qos}, and run on power-limited devices. This limitation has spurred new analysis and designs for energy efficiency. For example, bounds on the minimum energy-per-bit have been established in \cite{art4-5} and power allocations for energy efficiency of the relay channel have been derived in \cite{art6-2,art5-4}.
Despite of this interest, energy efficiency analysis is mostly based on power allocations optimized for rate as in \cite{art5-4}. Coding and optimization directly for energy remain scarce. One reason is that the two problems of maximizing rate at a given network power and minimizing network power at a given rate are often thought to be  equivalent. Nevertheless, neither optimization problem is fully characterized. Power allocations optimized for rate remain implicit without closed-form solutions, while power allocations for energy remain scarce and are mainly based on unrealistic constraints, such as full-duplex transmissions \cite{art6-2} or no individual power constraints \cite{art5-4,art6-2}.

In this paper, we explore energy optimization for the three-node relay channel in Gaussian noise environments, considering Gaussian signaling as typically used in such analysis to define upper-bounds for the system performance \cite{tse2005}.
We first propose a comprehensive half-duplex coding scheme based on time division for fading channels with Gaussian noise. We then explicitly optimize its resource allocation for energy efficiency at a desired source rate and individual node power constraints. We consider three objectives: the network, the relay and the source energy consumption. We next combine the proposed schemes into a generalized one that maximizes the energy efficiency of the Gaussian relay channel and analyze its performance as a function of both the source rate and the mobile user location. The contributions are threefold: 
\begin{itemize}
\item We show that maximizing the source rate and minimizing the network energy consumption are not equivalent as often believed. Specifically, there are different power-optimized regimes that cannot be uncovered by maximizing rates. 
\item We highlight the different regimes of coding technique and optimal power allocation (full or partial decode-forward, with or without beamforming), which constitute the lower bound on the capacity of the Gaussian relay channel by using decode-forward.
\item We provide the optimal power allocation in closed-form.
\end{itemize}

The paper is organized as follows. The new half-duplex coding scheme is analyzed for the Gaussian relay channel in Section \ref{sec:gaussian_model}. We optimize the resource allocation for network, relay and source energy efficiency in Section \ref{sec:X-EE} and build the generalized scheme in Section \ref{sec:G-EE}. Section \ref{sec:simulations} presents the performance analysis and Section \ref{sec:conclusion} concludes this paper.

%%%%%%%%%%%%%%%%%%%%%%%%%%%%%%%%%%%%%%%%%%%%%%%%%%%%%%%%%%%%%%%%%%%%%%%%%%%%%%%
\begin{figure*}[!t]
% ensure that we have normalsize text
\normalsize
% Store the current equation number.
\setcounter{MYtempeqncnt}{\value{equation}}
% Set the equation number to one less than the one
% desired for the first equation here.
% The value here will have to changed if equations
% are added or removed prior to the place these
% equations are referenced in the main text.
\setcounter{equation}{3}
\small{
\begin{equation}\label{eq:gauss_const}
\begin{split}
R \leq & \; \theta \log_2 \left( 1+ \frac{\left( \eta_1+\rho_1 \right)P_s \vert h_d \vert ^2 }{N}\right) + \bar{\theta} \log_2 \left( 1+ \frac{\left( \eta_2+\rho_2 \right)P_s \vert h_d \vert ^2 + \rho_r P_r \vert h_r \vert ^2 + 2 \sqrt{P_s \vert h_d \vert ^2 P_r \vert h_r \vert ^2 \rho_2 \rho_r}}{N}\right) = I_1 \\
\hspace*{-20pt} R \leq & \; \theta \log_2 \left( 1+ \frac{\rho_1 P_s \vert h_s \vert ^2 }{N + \eta_1 P_s \vert h_s \vert ^2}\right) + \theta \log_2 \left( 1+ \frac{ \eta_1 P_s \vert h_d \vert ^2 }{N}\right) + \bar{\theta} \log_2 \left( 1+ \frac{ \eta_2 P_s \vert h_d \vert ^2 }{N}\right) = I_2
\end{split}
\end{equation}
}
% The spacer can be tweaked to stop underfull vboxes.
%\vspace*{2pt}
% IEEE uses as a separator
\hrulefill
% Restore the current equation number.
\setcounter{equation}{\value{MYtempeqncnt}}
\end{figure*}

%%%%%%%%%%%%%%%%%%%%%%%%%%%%%%%%%%%%%%%%%%%%%%%%%%%%%%%%%%%%%%%%%%%%%%%%%%%%%%%

\section{A Comprehensive Half-Duplex Decode-Forward Scheme for Gaussian Channels}
\label{sec:gaussian_model}

\subsection{Channel model}

We consider a half-duplex channel with time division, such that the transmission is carried out in two phases within each code block of normalized length. During the first phase, of duration $\theta \in [0,1]$, the source transmits while the relay listens. During the second phase, of duration $\bar{\theta}  = \left( 1- \theta \right)$, both the source and the relay transmit. % Considering this time division, the channel during the first phase is $p(y,y_r \vert x)$, and during the second is $p(y \vert x,x_r)$. %\vspace*{-10pt}
We consider complex Gaussian channels, where $h_d$, $h_s$ and $h_r$ respectively stand for the gain of the direct link, the source-to-relay link and the relay-to-destination link, as depicted in Figure \ref{coding}. We assume that channel gain information is known globally at all nodes and consider independent Gaussian noises $Z_1$, $Z_2$ and $Z_r$ with variance $N$. The half-duplex Gaussian relay channel can be written as follows
\begin{equation}  \label{channel}
\begin{split}
\textit{Phase 1:} & \quad Y_r = h_s X_1 + Z_r \quad ; \quad Y_1 = h_d X_1 + Z_1 \\% \quad ; \quad
\textit{Phase 2:} & \quad Y_2 = h_d X_2 + h_r X_r + Z_2
\end{split}
\end{equation}

\noindent where at each time either $Y_r$ and $Y_1$ occur together or only $Y_2$ can occur.

\subsection{Coding scheme for the Gaussian relay channel}

We consider Gaussian signaling and superposition coding as in the half-duplex partial decode-forward scheme presented in our conference version \cite{ITW2011} for the discrete memoryless and Gaussian channels. For cooperative transmission, superposition modulation benefits from a renewed interest for potential implementation. As shown in \cite{Myths_and_facts}, this strategy outperforms other techniques, such as bit-interleaved coded modulation, at even lower receiver complexity. 

\begin{figure}
\centering \includegraphics[width = 0.98\columnwidth]{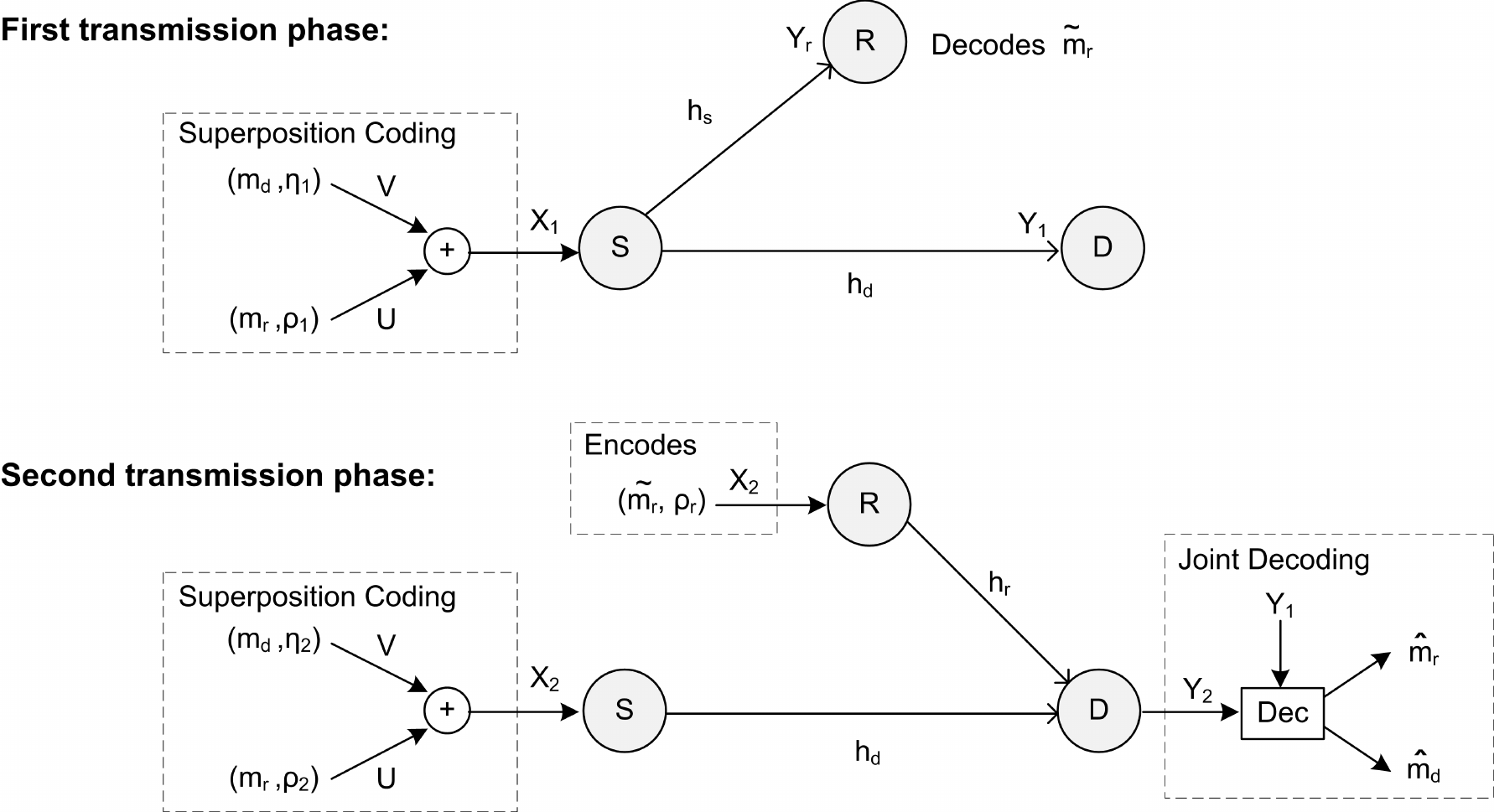} 
\caption{A Half-Duplex Coding Scheme for Relay Channels}
\label{coding}
\end{figure}

We describe the coding scheme as follows.
To send a message $m$ of rate $R$ to the destination, the source performs message splitting. It divides the initial message into two parts $(m_d,m_r)$, with rates $R_d$ and $R_r$ respectively, where $R_d+R_r=R$. Both are encoded at the source using superposition coding. The message $m_d$ is directly decoded by the destination at the end of the second phase, whereas $m_r$ is intended to be relayed. This scheme is depicted in Figure \ref{coding}. 

The source and the relay have individual power constraints $P_s$ and $P_r$ within the same bandwidth. At each phase, each node allocates to each message a portion of its available transmit power.
Denote $\eta_1$ and $\eta_2$ as the portion of source power $P_s$ allocated to $m_d$ in the first and second phase respectively, similarly, $\rho_1$ and $\rho_2$ as the portion for $m_r$.
Denote $\rho_r$ as the portion of the relay power $P_r$ used to forward $\tilde{m}_r$ to the destination. We consider transmit power constraint at each node such that
\begin{equation} 
\begin{split}
P_s^{(c)}& = \theta \left( \eta_1+\rho_1 \right) P_s + \bar{\theta}\left( \eta_2+\rho_2\right)P_s \leq \; P_s  
\\
P_r^{(c)}& = \bar{\theta} \rho_r P_r \leq \; P_r \, ,
\end{split}\label{eq:power_const}
\end{equation}
where superscript $(c)$ refers to the consumed power. 

Applying the scheme to the Gaussian relay channel, the transmit signal can be written in two separate phases as the channel in \eqref{channel} as follows
\begin{align}\label{eq:gaussian_channel_model}
%\begin{split}
\textit{Phase 1:} \quad X_1 &= \sqrt{\rho_1 P_s} U + \sqrt{\eta_1 P_s} V
 \\%\quad ;\quad
 \nonumber
 \textit{Phase 2:}
\quad X_r & = \sqrt{\rho_r P_r} U \quad ;\quad
X_2 = \sqrt{\rho_2 P_s} U + \sqrt{\eta_2 P_s} V 
%\end{split}
\end{align}
where $U \left( m_r \right)\sim \mathcal{N}(0,1)$ and $V \left( m_d \right)\sim \mathcal{N}(0,1)$ are independent Gaussian signals that carry messages $m_r$ and $m_d$ respectively. Note that $U$ appears in both $X_r$ and $X$, which allows a beamforming gain at the destination.

Joint decoding is considered at the receivers. At the end of Phase 1, the relay decodes $m_r$, i.e. chooses the unique $\tilde{m}_r$ for which $U$ and $Y_r$ are typical. At the end of Phase 2, the destination jointly decodes $(m_d,m_r)$ by choosing the unique $(\hat{m}_d,\hat{m}_r)$ for which $U$ and $X_1$ are jointly typical with $Y_1$, and simultaneously $U$, $X_2$ and $X_r$ are jointly typical with $Y_2$. More details on decoding can be found in our previous work \cite{ITW2011}. Note that instead of joint typicality, maximum likelihood decoding can be used to achieve the same rate.

For Gaussian relay channels, all rates satisfying constraints in Eq.\eqref{eq:gauss_const} at the top of the page are achievable.
%%%%%%%%%%%%%%%%%%%%%%
\addtocounter{equation}{1}
%%%%%%%%%%%%%%%%%%%%%%
In this equation, $\theta \in [0,1]$, and power set $(\eta_1, \rho_1, \eta_2, \rho_2, \rho_r)$ is non-negative and satisfies constraint \eqref{eq:power_const}.
%\label{Archievability_Gaussian}
Both rate constraints are results of the decoding at the relay over the first phase and joint decoding at the destination simultaneously over both phases. The first constraint captures the coherent transmission between source and relay in the second phase, whereas the second constraint comes from the decoding of $m_r$ at the relay and decoding of $m_d$ at the destination. 
We refer to \cite{ITW2011} for the general proof in the discrete memoryless channel case. 
This coding scheme includes as special cases direct transmissions, two-hops relaying, as well as the maximum-rate scheme proposed in \cite{Host-Madsen,Host-MadsenJournal}.

\subsection{Energy-optimization problem}

We propose to optimize the above coding scheme for energy efficiency in the Gaussian relay channel at a given source rate. We consider three cases: the network, the relay and the source power consumptions. Specifically, we look for the optimal set of power allocation $(\rho_1,\eta_1,\rho_2,\eta_2,\rho_r)$ solving the following general problem:
\begin{align}
 \min  \quad &   \omega_s \left[ \theta \left( \eta_1+\rho_1 \right) P_s + \bar{\theta}\left( \eta_2+\rho_2\right)P_s \right]+ \omega_r\bar{\theta}\rho_r P_r \label{eq:min_energy_pbm} \\
\text{s.t. } \quad &  \begin{array}{l l}
I_1 \geq R  & ; \quad I_2 \geq R \\
P_s^{(c)}\leq P_s  & ; \quad P_r^{(c)}\leq P_r
\end{array}
\nonumber % \\ \nonumber 
\end{align}
where $\omega_s$ and $\omega_r$ are either 0 or 1, depending on the targeted optimization. The two rate constraints ensure achievability of the source rate, given two individual power constraints.

Optimization for Gaussian signaling in Gaussian channel as formulated in \eqref{eq:min_energy_pbm} provides an upper-bound on the performance of practical systems. It defines the limit of what is possible in practice, as well as gives an insight of how a practical scheme should be structured. Especially, it is found in  \cite{SendonarisPartI} that the system that emulates most closely the structure of signals defined by information theory also reaches the highest performance.
The authors of \cite{SendonarisPartI,SendonarisPartII} analyze the maximum rates allowed by information theory for the multiple-access Gaussian channel using cooperation, as well as implementation in a CDMA system. 
Other analysis is provided in \cite{zhou2012} for the interference Gaussian channel in a tri-sectored OFDMA network.
In this reference, an interference mitigation scheme with power allocation is proposed for Gaussian signaling and then is simulated for various practical modulations and coding rates. The practical system approaches the channel capacity with Gaussian signaling within 1.2 bit per channel use.
These examples of existing works illustrate that optimization for Gaussian signaling is a valid approach for determining optimal power allocation for practical implementation.

\section{Three energy efficient schemes \\ for the relay channel}
\label{sec:X-EE}

In this section, we explore energy optimization for the relay channel considering three objectives: the network, the relay and the source energy consumption. Then, we compare the maximum achievable rate of each optimized scheme. %\vspace*{-10pt}

%%%%%%%%%%%%%%%%%%%%%%%%%%%%%%%%%%%%%%%%%%%%%%%%%%%%%%%%%%
\subsection{Network energy optimal set of power allocation (N-EE)}
\label{sec:N-EE}

In this optimization, we consider the total source and relay power consumption during both transmission phases.  To minimize the total consumption while maintaining a desired source rate, we analyze the following problem ($\omega_s = 1$, $\omega_r = 1$), denoted as N-EE: %\vspace*{-10pt}
\begin{align}
 \min  \quad &   \left[ \theta \left( \eta_1+\rho_1 \right) P_s + \bar{\theta}\left( \eta_2+\rho_2\right)P_s \right]+ \bar{\theta}\rho_r P_r \label{eq:min_energy_pbm_N} \\
\text{s.t. } \quad &  \begin{array}{l l}
I_1 \geq R  & ; \quad I_2 \geq R \\
P_s^{(c)}\leq P_s  & ; \quad P_r^{(c)}\leq P_r 
\end{array} \nonumber
\end{align}

The optimal power allocation depends on the desired rate. If the source rate is higher than a certain threshold $R^{(n)}$, then partial decode-forward is applied (sub-scheme $A^{(n)}$), as detailed in Proposition \ref{prop:A_n}. Otherwise, the source does not split its message and uses full decode-forward (sub-scheme $B^{(n)}$), with power allocation as given in Proposition \ref{prop:B_n}. 
Note that the power consumption of N-EE is a continuous function of the source rate, even at $R^{(n)}$. This result is summarized in Algorithm \ref{Algo_N_EE} and the proof can be found in Appendix \ref{appendix:min_network}.

\begin{table}
%\begin{algorithm}
%
%%\begin{algorithm}
\centering \fbox{
\begin{minipage}{0.9\columnwidth}
\begin{algorithmic}
	\IF{$\vert h_s \vert ^2 \geq \vert h_d \vert ^2$}
	\STATE Find $\rho_1^\star \in \left[0, 1/\theta\right]$ satisfying $g_1(\rho_1^\star) = 0$ where
\begin{align}
& g_1(\rho_1^\star) = v_d^{1/\bar{\theta}} - 1
	+ \frac{\vert h_r \vert ^2}{\vert h_d \vert ^2}  \left( \frac{\vert h_s \vert ^2}{\vert h_d \vert ^2}
		v_s^{1/\bar{\theta}} -1 \right) \label{eq:solve_rho_1n} \\
& \text{with } \quad v_s= \frac{2^{R}}{\left( 1+ \frac{\rho_1^\star P_s \vert h_s \vert ^2}{N}\right)} 
\nonumber \\%\quad
& \text{and } \quad v_d= \frac{2^{R}}{\left( 1+ \frac{\rho_1^\star P_s \vert h_d \vert ^2}{N}\right)} = \frac{2^R \frac{\vert h_s \vert ^2}{\vert h_d \vert ^2} v_s}{2^R + \left(\frac{\vert h_s \vert ^2}{\vert h_d \vert ^2}-1\right)v_s}
\nonumber
\end{align}
\IF{$\rho_1^\star$ exists}
	\STATE $R^{(n)} \leftarrow \theta \log_2 \left(1+ \frac{\rho_1^\star P_s \vert h_s \vert^2}{N} \right) $
\ELSE
	\STATE $R^{(n)} \leftarrow \infty $
\ENDIF
	\IF{$R \geq R^{(n)}$}
		\STATE Use sub-scheme $A^{(n)}$ (partial DF) given in Proposition \ref{prop:A_n};
	\ELSE
		\STATE Use sub-scheme $B^{(n)}$ (full DF) given in Proposition \ref{prop:B_n};
	\ENDIF
\ELSE
	\STATE Use direct transmission or declare outage if not feasible
\ENDIF
%\ENSURE The power constraints \eqref{eq:power_const} are met. Otherwise, declare outage.
\end{algorithmic}
\end{minipage}
}
\caption{Optimal scheme for network energy efficiency, N-EE, \quad solving \eqref{eq:min_energy_pbm_N} \vspace*{-20pt}}
%\end{algo}
%\end{algorithm}
\label{Algo_N_EE}
\end{table}

\begin{prop}
In sub-scheme $A^{(n)}$, the source splits its message into two parts $m_r$ and $m_d$. \\
\hspace*{15pt}$\bullet$ In Phase 1, the source sends $m_r$ with power $\rho_1^\star P_s$ and rate $R_r=\theta \log_2 \left( 1+ \frac{\rho_1 P_s \vert h_s \vert ^2 }{N}\right)$. The relay decodes as $\tilde{m}_r$.\\
\hspace*{15pt}$\bullet$ In Phase 2, the relay sends $\tilde{m}_r$ with power $\rho_r^\star P_r$ and the source sends $(m_r,m_d)$ with power $(\rho_2^\star P_s,\eta_2^\star P_s)$. The rate of $m_d$ is $R_d= R-R_r$. \\
To compute the optimal power allocation set for $A^{(n)}$, $\rho_1^\star$ is first found numerically by solving Eq. \eqref{eq:solve_rho_1n} of Algorithm \ref{Algo_N_EE}. % where
Next, $(\rho_r^\star,\rho_2^\star,\eta_1^\star,\eta_2^\star)$ are deduced from $\rho_1^\star$ as follows:

\noindent \small
\begin{align*}
\eta_1^\star & = 0 \; ; \quad
\rho_r^\star   =  \frac{N \vert h_r \vert ^2}{P_r \left(\vert h_r \vert ^2 + \vert h_d \vert ^2 \right)^2} \left(\frac{2^{R/ \bar{\theta}}}{\left( 1+\frac{\rho_1^\star P_s \vert h_d \vert ^2}{N}\right)^{\theta / \bar{\theta}}}
\right. \nonumber \\
& \qquad \qquad \qquad  \left. 
-\frac{2^{R/ \bar{\theta}}}{\left( 1+ \frac{\rho_1^\star P_s \vert h_s \vert ^2 }{N} \right)^{\theta / \bar{\theta}}} \right)  \\
%\\ %
\eta_2^\star &=  \left(\frac{2^{R/ \bar{\theta}}}{\left( 1+ \frac{\rho_1^\star P_s \vert h_s \vert ^2 }{N} \right)^{\theta / \bar{\theta}}} -1\right) \frac{N}{P_s \vert h_d \vert ^2}  \; ; \quad %\\ %
\rho_2^\star = \frac{P_r \vert h_d \vert^2}{P_s \vert h_r \vert^2} \rho_r^\star 
\end{align*}
\label{prop:A_n} \end{prop} 
\normalsize

Regarding the computation of optimal allocation, solving Eq. \eqref{eq:solve_rho_1n} of Algorithm \ref{Algo_N_EE} in terms of $v_s$, rather than in terms of $\rho_1^\star$, reduces to solving a polynomial equation and is sufficient to deduce the whole allocation. This shows that computing the optimal allocation is relatively simple, especially in case of equal time division ($\theta = \bar{\theta} = \frac{1}{2}$) which is widely used in practical systems.

As shown in Algorithm \ref{Algo_N_EE}, partial decode-forward is applied as long as $R^{(n)} \leq R \leq R^{(n)}_{\max}$. The lower bound $R^{(n)}$ comes from the non-negativity constraint on $\eta_2^\star$. The upper bound $ R^{(n)}_{\max}$ corresponds to the maximum rate that is achievable given the power constraints defined in \eqref{eq:power_const}. 
We will come back to this upper bound in Section \ref{prop:scheme_comp}. 
Note that, when the direct link is close to 0, the maximum feasible rate for $m_d$ goes to zero. Due to power constraints, outage can occur at some rate $R \leq R^{(n)}$, such that sub-scheme $A^{(n)}$ may not exist. 
When $R \leq R^{(n)}$, we apply sub-scheme $B^{(n)}$ as discussed next.

\begin{prop}
In sub-scheme $B^{(n)}$, the source does not split its message $m$.\\
\hspace*{15pt}$\bullet$ In Phase 1, the source sends $m$ with power $\rho_1^\dagger P_s$ and the relay decodes as $\tilde{m}$. \\
\hspace*{15pt}$\bullet$ In Phase 2, the relay sends $\tilde{m}$ with power $\rho_r^\dagger P_r$ and the source sends $m$ with power $\rho_2^\dagger P_s$. \\
The optimal power allocation for sub-scheme $B^{(n)}$ is 

\noindent \small
\begin{align*}
	\eta_1^\dagger & = \eta_2^\dagger = 0 \\
    \rho_1^\dagger & =  \left(2^{R/ \theta}-1 \right) \frac{N}{P_s \vert h_s \vert ^2}  \; ; \quad 
    \rho_2^\dagger = \frac{P_r \vert h_d \vert^2}{P_s \vert h_r \vert^2} \rho_r^\dagger   
    \\
    \rho_r^\dagger & = \frac{N \vert h_r \vert ^2}{P_r \left(\vert h_r \vert ^2 + \vert h_d \vert ^2 \right)^2} \left( \frac{2^{R/ \bar{\theta}}}{\left( 1+\frac{\rho_1^\dagger P_s \vert h_d \vert ^2}{N}\right)^{\theta / \bar{\theta}}} -1\right).
\end{align*} \label{prop:B_n} \end{prop}
\normalsize

\noindent $B^{(n)}$ is applied when the source rate is low. In this case, full decode-forward is energy-efficient and the whole message benefits from the beamforming gain between the source and the relay.

\begin{figure}
\begin{center}
\includegraphics[width=0.65\columnwidth]{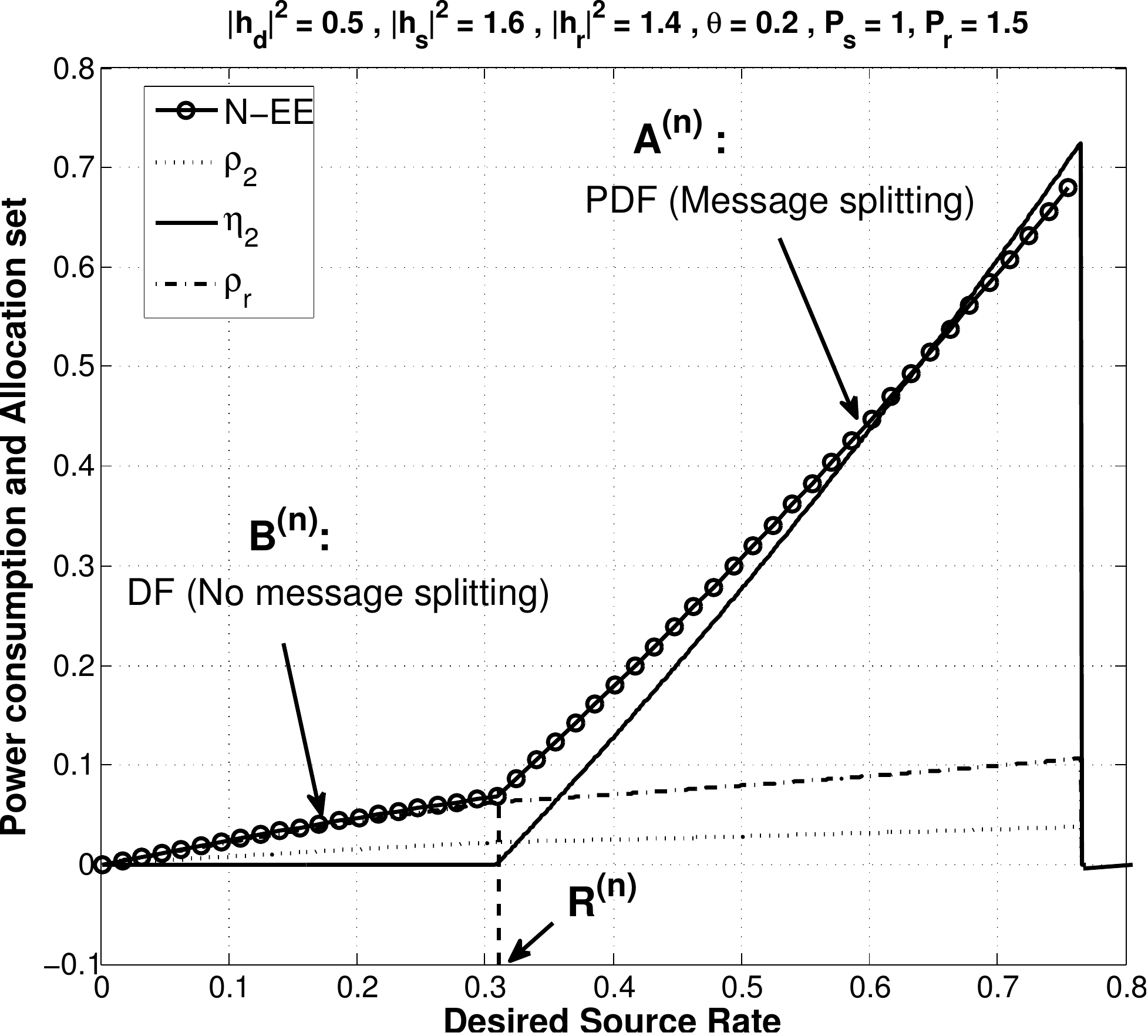}
\end{center}
\caption{Power consumption for N-EE and allocation set during phase 2.}
\label{Fig:Example_subschemes_network}
\end{figure}

%\vspace{15pt}
As an example, Figure \ref{Fig:Example_subschemes_network} illustrates the power consumptions of the above two sub-schemes, which differ mainly in the second phase. Thus, the figure plots the set of optimal power allocation and the total energy consumed by the source and the relay during this phase, as a function of the source rate. When the source rate is not achievable given power constraints, an outage occurs, which is depicted by a cut-off in the curve.

%%%%%%%%%%%%%%%%%%%%%%%%%%%%%%%%%%%%%%%%%%%%%%%%%%%%%%%
\subsection{Relay energy optimal set of power allocation (R-EE)}
\label{sec:R-EE}

Next, we minimize the relay power consumption only and allow the source to consume up to $P_s$ during both phases of the transmission. This scheme is of particular interest in networks where the relay is shared and has to serve many users, or where the relay has its own data to send. Thus, we analyze the following optimization problem, denoted as R-EE: %\vspace*{-20pt}
\begin{align}
 \min  \quad &   \bar{\theta}\rho_r P_r  \label{eq:min_energy_pbm_R} \\
\text{s.t. } \quad &  \begin{array}{l l}
I_1 \geq R  & ; \quad I_2 \geq R \\
P_s^{(c)}\leq P_s  & ; \quad P_r^{(c)}\leq P_r
\end{array} \nonumber
%\\
%\text{where } \quad &  R_r=\theta \log_2 \left( 1+ \frac{\rho_1 P_s \vert h_s \vert ^2 }{N}\right) \quad \text{and} \quad R_d=R-R_s
%\nonumber % \\ \nonumber 
\end{align}

%\vspace*{10pt}
Once again, the optimal power allocation depends on the desired rate. Direct transmission (sub-scheme $C^{(r)}$) is used whenever feasible, i.e. as long as the source rate is under the capacity of the direct link, denoted as $R_1^{(r)}$. Note that direct transmissions can be optimal for relay energy but not for source or network energy. Indeed, the optimization problem considered here is relaxed in terms of the source consumption. This implies that some sub-schemes can be optimal for R-EE but not for N-EE.

If direct transmission is not feasible, the optimal scheme for relay energy uses, similarly to N-EE, either partial decode-forward (sub-scheme $A^{(r)}$) or full decode-forward (sub-scheme $B^{(r)}$), depending on a certain threshold $R_2^{(r)}$. This result is summarized in Algorithm \ref{Algo_R_EE} and the proof can be found in Appendix \ref{appendix:min_relay}.

\begin{table}
\centering \fbox{
\begin{minipage}{0.9\columnwidth}
\begin{algorithmic}
\STATE $R^{(r)}_1 \leftarrow \log_2 \left( 1+ \frac{P_s \vert h_d \vert ^2}{N}\right)$
\IF{$R \leq R^{(r)}_1$}
	\STATE Use direct transmission, called sub-scheme $C^{(r)}$, as given in Proposition \ref{prop:C_r};
	\ELSIF {$\vert h_s \vert ^2 \geq \vert h_d \vert ^2$}
		\STATE Find $\rho_1^\star \in \left[0, 1/\theta\right]$ satisfying $g_2(\rho_1^\star) = 0$ where
\begin{align}
g_2(s) & =  \frac{2^{R/ \bar{\theta}}}{\left( 1+\frac{s P_s \vert h_d \vert ^2}{N}\right)^{1 / \bar{\theta}}}
	- 1
	\label{eq:solve_rho_1r}
\\
	& 
	+ g_3(s) \left( \frac{\vert h_s \vert ^2}{\vert h_d \vert ^2} \frac{2^{R/ \bar{\theta}}}{\left( 1+ \frac{s P_s \vert h_s \vert ^2 }{N} \right)^{1 / \bar{\theta}}} -1 \right)
  \nonumber\\
\text{and} \quad g_3(s) & =  \frac{2^{R/ \bar{\theta}}}{\left( 1+\frac{s P_s \vert h_d \vert ^2}{ N}\right)^{\theta / \bar{\theta}}}- \frac{2^{R/ \bar{\theta}}}{\left( 1+ \frac{s P_s \vert h_s \vert ^2 }{ N} \right)^{\theta / \bar{\theta}}} \label{h_2}
\end{align}
	\IF{$\rho_1^\star$ exists}
	\STATE $R^{(r)}_2 \leftarrow \theta \log_2 \left(1+ \frac{\rho_1^\star P_s \vert h_s \vert^2}{N} \right) \label{R_2}$
\ELSE
	\STATE $R^{(r)}_2 \leftarrow \infty $
\ENDIF
		
		\IF{$R \geq R^{(r)}_2$}%
			\STATE Use sub-scheme $A^{(r)}$ (partial DF) given in Proposition \ref{prop:A_r};
		\ELSE	
		\STATE Use sub-scheme $B^{(r)}$ (full DF) given in Proposition \ref{prop:B_r};
		\ENDIF
	\ELSE
	\STATE Declare outage.
	\ENDIF
%\ENSURE The power constraints \eqref{eq:power_const} are met. Otherwise, declare outage.
\end{algorithmic}
\end{minipage}
}
\caption{Optimal scheme for relay energy efficiency, R-EE, solving \eqref{eq:min_energy_pbm_R} \vspace*{-20pt}}
\label{Algo_R_EE} %\vspace*{-15pt}
\end{table}

In the following propositions, we present each sub-scheme and associated power allocation going from low to high rates.
A scheme which aims at minimizing the relay energy consumption should always give priority to direct transmissions, such that, if possible, the relay is not used at all. Thus, as long as the direct link is strong enough to support the source rate, R-EE uses sub-scheme $C^{(r)}$. %, which corresponds to direct transmission.
\begin{prop}
Sub-scheme $C^{(r)}$ is direct transmission.
The optimal power allocation is as follows.
\begin{align*}
    \eta_1^\ddagger & = \eta_2^\ddagger = \left(2^{R}-1 \right) \frac{N}{P_s \vert h_d \vert ^2} \; ; \quad 
    \rho_1^\ddagger = \rho_2^\ddagger = \rho_r^\ddagger = 0 .
\end{align*}
\label{prop:C_r}
\end{prop}

%\vspace{-15pt}
If the source rate increases above the capacity of the direct link, relaying is required and either sub-scheme $A^{(r)}$ or $B^{(r)}$ is applied, as discussed next. These sub-schemes are similar to those in N-EE but with different power allocation. In the case of R-EE, the source consumes all its available power.
%: $\theta \left( \eta_1+\rho_1 \right) P_s + \bar{\theta}\left( \eta_2+\rho_2\right)P_s = P_s $
If $R \leq R^{(r)}_2$ as in Algorithm \ref{Algo_R_EE}, we apply sub-scheme $B^{(r)}$.
\begin{prop}
In sub-scheme $B^{(r)}$, the source uses full decode-forward.\\
\hspace*{15pt}$\bullet$ In Phase 1, the source sends $m$ with power $\rho_1^\dagger P_s$. The relay decodes as $\tilde{m}$. \\
\hspace*{15pt}$\bullet$ In Phase 2, the relay sends $\tilde{m}$ with power $\rho_r^\dagger P_r$ and the source sends $m$ again, with power $\rho_2^\dagger P_s$. \\
The optimal power allocation for sub-scheme $B^{(r)}$ is
\begin{align*}
     \eta_1^\dagger &= \eta_2^\dagger = 0 \\
    \rho_1^\dagger & = \left( 2^{R / \theta }-1 \right) \frac{N}{P_s \vert h_s \vert ^2} \; ; \quad
    \rho_2^\dagger = \frac{1-\rho_1^\dagger \theta }{\bar{\theta}}
    \; ; \quad
%    \end{align*} \vspace{-30pt}
%    \begin{align*}
	\\
    \rho_r^\dagger  &=  \frac{N}{P_r \vert h_r \vert ^2} \left( \sqrt{g_3(\rho_1^\dagger)} - \sqrt{\frac{P_s \vert h_d \vert^2 \rho_2^\dagger}{N}}\right)^2 . 
\end{align*} \label{prop:B_r} \end{prop} 
%\vspace*{-10pt}
Note that sub-scheme $B^{(r)}$ only exists if the channel gains are such that $R^{(r)}_1 < R^{(r)}_2$. This condition essentially depends on the strength of the direct link. When $R^{(r)}_1 \geq R^{(r)}_2$, sub-scheme $A^{(r)}$ is directly applied.

\begin{figure}
\begin{center}
\includegraphics[width=0.65\columnwidth]{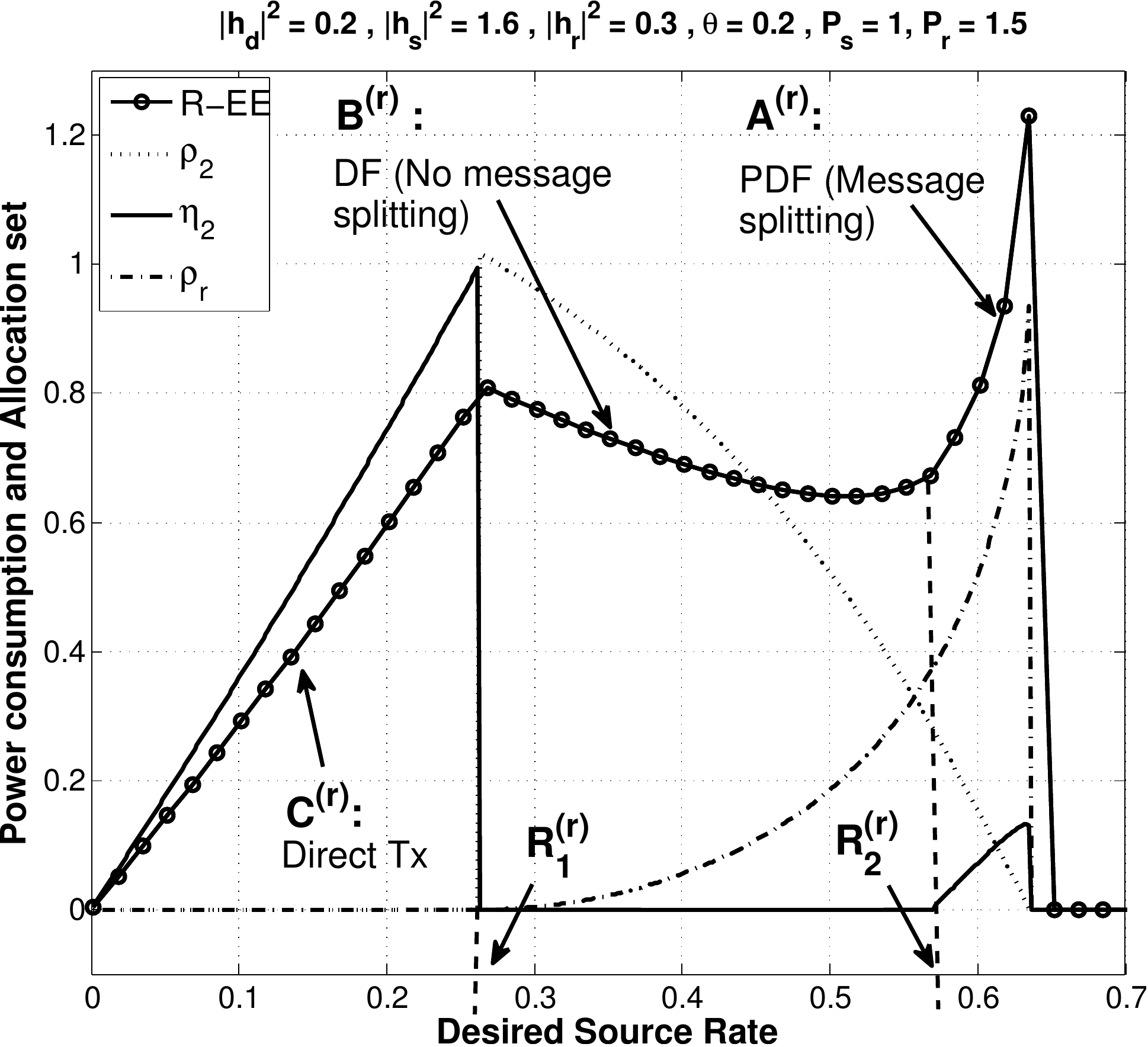}
\end{center}
\caption{Power consumption for R-EE and allocation set during phase 2.}
\label{Fig:Example_subschemes_relay}
\end{figure}

When the source rate increases even further to be above $R^{(r)}_2$, message splitting is required and sub-scheme $A^{(r)}$ is applied.
\begin{prop}
In sub-scheme $A^{(r)}$, the source splits its message into two parts $m_r$ and $m_d$.  \\
\hspace*{15pt}$\bullet$ In Phase 1, the source sends $m_r$ with power $\rho_1^\star P_s$ and rate $R_r=\theta \log_2 \left( 1+ \frac{\rho_1 P_s \vert h_s \vert ^2 }{N}\right)$. The relay decodes as $\tilde{m}_r$ \\
\hspace*{15pt}$\bullet$ In Phase 2, the relay sends $\tilde{m}_r$ with power $\rho_r^\star P_r$ and the source sends $(m_r,m_d)$ with power $(\rho_2^\star P_s,\eta_2^\star P_s)$. The rate of $m_d$ is $R_d= R-R_r$.\\
To compute the optimal power allocation set for sub-scheme $A^{(r)}$, $\rho_1^\star$ is first found numerically by solving Eq. \eqref{eq:solve_rho_1r} of Algorithm \ref{Algo_R_EE}. %where
Next, $g_3(\rho_1^\star)$ is computed using \eqref{h_2} and $(\rho_r^\star,\rho_2^\star,\eta_1^\star,\eta_2^\star)$ are deduced from $\rho_1^\star$ and $g_3(\rho_1^\star)$ as follows: %\vspace*{-10pt}
\begin{align*}
\eta_2^\star & =  \left(\frac{2^{R/ \bar{\theta}}}{\left( 1+ \frac{\rho_1^\star P_s \vert h_s \vert ^2 }{N} \right)^{\theta / \bar{\theta}}} -1\right) \frac{N}{P_s \vert h_d \vert ^2} 
\\
\eta_1^\star & = 0
 \; ; \quad
\rho_2^\star = \frac{\left(1-\rho_1^\star \theta \right)}{\bar{\theta}} - \eta_2^\star
\\
\rho_r^\star  &=  \frac{N}{P_r \vert h_r \vert ^2} \left(
\sqrt{g_3(\rho_1^\star) } - \sqrt{\frac{P_s \vert h_d \vert^2 \rho_2^\star}{N}}
%\right. \\
%& \left. \quad \quad \quad \quad
\right)^2.
\end{align*}
%and $\rho_1^\star$ is found numerically by solving $g_2(\rho_1^\star) = 0$ where
\label{prop:A_r}
\end{prop}
%\vspace*{-20pt}
If $\eta_i^\star$ and $\rho_i^\star$ as in Proposition \ref{prop:A_r} do not satisfy the power constraints in \eqref{eq:power_const}, then the desired source rate cannot be achieved and outage is declared. For example and similarly to $A^{(n)}$, when the direct link is very weak, sub-scheme $A^{(r)}$ may not exist.

Figure \ref{Fig:Example_subschemes_relay} illustrates the above sub-schemes and plots the total energy consumption during the second phase as a function of the source rate. Somewhat unexpectedly, the energy consumed during this second phase is decreasing in R when applying sub-scheme $B^{(r)}$. Since the destination decodes the message using the received signals during both phases, the optimization shows that it is relay energy-optimal for the source to allocate more energy during the first phase. Nevertheless, the total energy consumed during both phases is strictly increasing with rate, as expected. %\vspace*{-10pt}

%%%%%%%%%%%%%%%%%%%%%%%%%%%%%%%%%%%%%%%%%%%%%%%%%%%%%%%
\subsection{Source energy optimal set of power allocation (S-EE)}
\label{sec:S-EE}

As the last problem, we minimize the source power consumption alone, which is relevant to networks in which the relay is not power critical. We thus allow the relay to consume up to $P_r$ during both transmission phases. We analyze the following optimization problem, denoted as S-EE: %\vspace*{-5pt}
\begin{align}
 \min  \quad &   \theta \left( \eta_1+\rho_1 \right) P_s + \bar{\theta}\left( \eta_2+\rho_2\right)P_s  \label{eq:min_energy_pbm_S} \\
\text{s.t. } \quad &  \begin{array}{l l}
I_1 \geq R  & ; \quad I_2 \geq R \\
P_s^{(c)}\leq P_s  & ; \quad P_r^{(c)}\leq P_r
\end{array}
\nonumber % \\ \nonumber 
\end{align}

The optimal scheme for source energy efficiency is a function of the source rate and is composed of four sub-schemes: two-hop relaying ($D^{(s)}$), decode-forward ($B^{(s)}$), partial decode-forward with beamforming ($A^{(s)}$) and without beamforming($C^{(s)}$). Note that two-hop relaying and partial decode-forward without beamforming are optimal only for source energy but not for relay or network energy. These sub-schemes are applied depending on the source rate compared with two thresholds $R^{(s)}_1$ and $R^{(s)}_2$, which are closely related to the non-negativity constraints on $\rho_2$ and $\eta_2$. 
This result is summarized in Algorithm \ref{Algo_S_EE} and the proof can be found in Appendix \ref{appendix:min_source}. Note that the functions $g_2$ and $g_3$ of Algorithm \ref{Algo_S_EE} are the same in Algorithm \ref{Algo_R_EE}.

\begin{table}
\centering \fbox{
\begin{minipage}{0.9\columnwidth}
\begin{algorithmic}
	\IF{$\vert h_s \vert ^2 \geq \vert h_d \vert ^2$}
	\STATE \vspace{-10pt}\begin{align*}
	R_{B} \leftarrow &\theta \log_2 \left(1 +\frac{\vert h_d \vert ^2}{\vert h_s \vert ^2} \left( 2^{R/\theta}-1\right) \right) + \\
	& \bar{\theta} \log_2 \left( 1+ \frac{P_r \vert h_r \vert^2}{\bar{\theta} N}\right)\\
	R_{C} \leftarrow &\theta \log_2 \left( \frac{\vert h_s \vert^2}{\vert h_d \vert^2} \right)
	\end{align*}
%	\STATE $$
%
\STATE Find $\rho_1^\star  \in \left[0, 1/\theta\right]$ satisfying $g_2(\rho_1^\star) = 0$ where
\begin{align}
g_2(s) & =  \frac{2^{R/ \bar{\theta}}}{\left( 1+\frac{s P_s \vert h_d \vert ^2}{ N}\right)^{\theta / \bar{\theta}}}
	- 1  \label{eq:solve_rho_1s}
\\ \nonumber &
	 + g_3(s)\left( \frac{\vert h_s \vert ^2}{\vert h_d \vert ^2} \frac{2^{R/ \bar{\theta}}}{\left( 1+ \frac{s P_s \vert h_s \vert ^2 }{N} \right)^{\theta / \bar{\theta}}} -1 \right)
 \\
\text{and} \quad g_3(s) & =  \frac{2^{R/ \bar{\theta}}}{\left( 1+\frac{s P_s \vert h_d \vert ^2}{ N}\right)^{\theta / \bar{\theta}}}- \frac{2^{R/ \bar{\theta}}}{\left( 1+ \frac{s P_s \vert h_s \vert ^2 }{ N} \right)^{\theta / \bar{\theta}}}
\label{h_3}
\end{align}
%	\STATE $R^{(s)}_1 \leftarrow \theta \log_2 \left( 1+ \frac{\rho_1^\star P_s \vert h_d \vert ^2 }{N}\right) + \bar{\theta} \log_2 \left( 1+ \frac{\eta_2^\star P_s \vert h_d \vert ^2 + \rho_r^\star P_r \vert h_r \vert ^2}{N}\right) $
	\IF{$R_{B} \leq R_{C} $}	%\COMMENT{the RD-link is the bottleneck}
	\STATE $R^{(s)}_1 \leftarrow R_{B}$	
	
	\STATE $R^{(s)}_2 \leftarrow \theta \log_2 \left(1+ \frac{\rho_1^\star P_s \vert h_s \vert^2}{N} \right)$ or $\infty$ if $\rho_1^\star$ does not exist
	\ELSE %\COMMENT{the SR-link is the bottleneck}
	\STATE $R^{(s)}_1 \leftarrow R_{C}$	
	
	\STATE $R^{(s)}_2 \leftarrow \bar{\theta} \log_2 \left(\frac{P_r \vert h_r \vert^2}{\bar{\theta}N} \right)$ $- \bar{\theta} \log_2 \left(\frac{1}{\left( 1+\frac{\rho_1^\star P_s \vert h_d \vert ^2}{ N}\right)^{\theta / \bar{\theta}}} \right. $ $ \left.-\frac{1}{\left( 1+\frac{\rho_1^\star P_s \vert h_s \vert ^2}{ N}\right)^{\theta / \bar{\theta}}} \right)$  or $\infty$ if $\rho_1^\star$ does not exist
	\ENDIF
	\IF{$R \geq R^{(s)}_1$ and $R \geq R^{(s)}_2$}
		\STATE Use sub-scheme $A^{(s)}$ (partial decode-forward with beamforming) given in Proposition \ref{prop:A_s};
	\ELSIF {$R^{(s)}_1 \leq R \leq R^{(s)}_2$ and $R_{B} \leq R_{C} $}
		\STATE Use sub-scheme $B^{(s)}$ (full decode-forward with beamforming) given in Proposition \ref{prop:B_s};
	\ELSIF {$R^{(s)}_1 \leq R \leq R^{(s)}_2$ and $R_{B} \geq R_{C} $}
		\STATE Use sub-scheme $C^{(s)}$ (partial decode-forward without beamforming) given in Proposition \ref{prop:C_s};
	\ELSIF {$R \leq R^{(s)}_1$ and $R \leq R^{(s)}_2$}
		\STATE Use sub-scheme $D^{(s)}$ (full decode-forward without beamforming) given in Proposition \ref{prop:D_s};
	\ENDIF
\ELSE
	\STATE Use the direct link or declare outage if not feasible.
\ENDIF
%\ENSURE The power constraints \eqref{eq:power_const} are met.
\end{algorithmic}
\end{minipage}
}
\caption{Optimal scheme for source energy efficiency, S-EE, solving \eqref{eq:min_energy_pbm_S} \vspace*{-20pt}}
\label{Algo_S_EE} \vspace*{-15pt}
\end{table}

\begin{figure}
\begin{center}
\includegraphics[width=0.9\columnwidth]{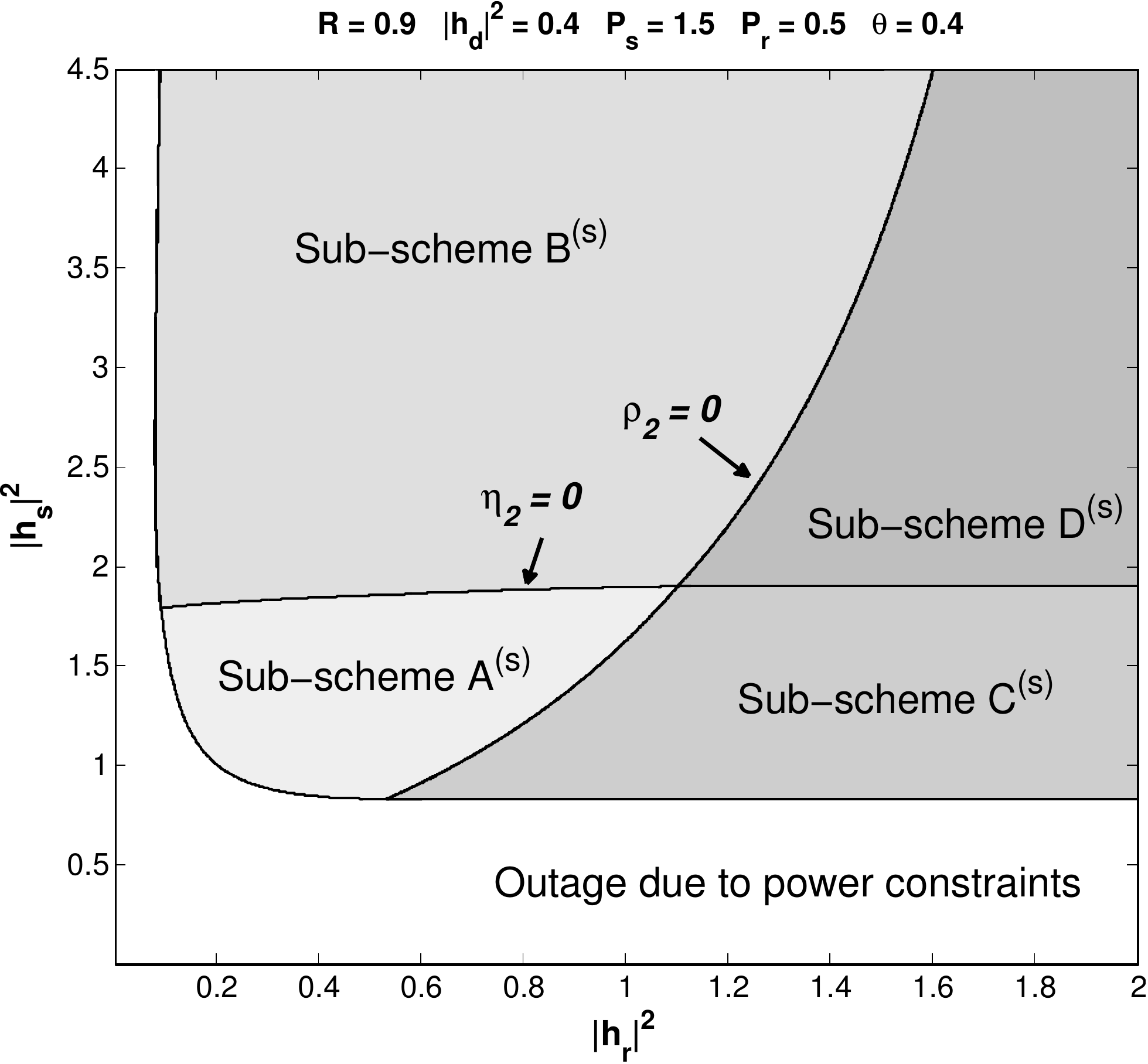}
\end{center}
\caption{Applied sub-schemes in S-EE, as function of the SR- and RD- links.}
\label{Fig:Description_subschemes_source}
\end{figure}

In the following propositions, we present each sub-scheme going from low to high rates and refer to Figure \ref{Fig:Description_subschemes_source} for an illustration. For very low rates, it is energy-efficient for the source to apply two-hop relaying, such that the source only consumes energy during the first phase. However, contrary to simple routing, the destination is listening during both transmission phases. This corresponds to sub-scheme $D^{(s)}$.
\begin{prop}
Sub-scheme $D^{(s)}$ is two-hop relaying with destination simultaneously decoding over both phases.  \\
\hspace*{15pt}$\bullet$ In Phase 1, the source sends $m$, the relay decodes as $\tilde{m}$. \\
\hspace*{15pt}$\bullet$ In Phase 2, the relay sends $\tilde{m}$ and the source is silent. \\
The optimal power allocation is as follows.
\begin{align*}
    \eta_1^\circ & = \eta_2^\circ = \rho_2^\circ = 0 \; ; \;
  	\rho_1 ^\circ = \left(2^{R/\theta}-1 \right) \frac{N}{P_s \vert h_s \vert ^2}
  	%\; ; \;
  	\\
    \rho_r^\circ &= \frac{N}{P_r \vert h_r \vert ^2} g_3(\rho_1^\circ)
\end{align*}
where $g_3$ is defined by \eqref{h_3} of Algorithm \ref{Algo_S_EE}.
\label{prop:D_s}
\end{prop}

If the rate increases, two-hop relaying is no longer energy-efficient for the source. In this case, either sub-scheme $C^{(s)}$ or $B^{(s)}$ is applied, depending on which of the SR- or RD-link is limiting.

If the SR-link is the bottleneck of the network, the second rate constraint $R \leq I_2$ is the limiting one and the constraint on the relay allocation is relaxed. In this case, $R_C \leq R_B$, as defined in Algorithm \ref{Algo_S_EE}, and sub-scheme $C^{(s)}$ is applied.
\begin{prop}
Sub-scheme $C^{(s)}$  uses partial decode-forward without beamforming.  \\
\hspace*{15pt}$\bullet$ In Phase 1, the source sends $m_r$ and the relay decodes as $\tilde{m}_r$.\\
\hspace*{15pt}$\bullet$ In Phase 2, the source sends $m_d$, the relay sends $\tilde{m_r}$. \\
The optimal power allocation for sub-scheme $C^{(s)}$ is such that
\begin{align*}
    \eta_1^\ddagger & = \rho_2^\ddagger  = 0 \; ; \quad 
    \rho_1^\ddagger  =  \frac{N}{P_s \vert h_s \vert ^2} \left( 2^R \left( \frac{\vert h_s \vert ^2}{\vert h_d \vert ^2}\right)^{\bar{\theta}} -1\right)
\\    
  %  \; ; \quad 
    \eta_2^\ddagger  &= \rho_1^\ddagger + \frac{N}{P_s}\left( \frac{1}{\vert h_s \vert^2}- \frac{1}{\vert h_d \vert^2} \right)
    %\frac{N}{P_s \vert h_d \vert ^2} \left( 
%    \frac{2^{R/ \bar{\theta}}}{\left( 1+\frac{\rho_1^\ddagger  P_s \vert h_s \vert ^2}{N}\right)^{\theta / \bar{\theta}}}- 1
%\right)
%\\
\; ; \quad 
    \rho_r^\ddagger =   \frac{N}{P_r \vert h_r \vert ^2} g_3(\rho_1^\ddagger)
%     \left( 
%    \frac{2^{R/ \bar{\theta}}}{\left( 1+\frac{\rho_1^\ddagger  P_s \vert h_d \vert ^2}{N}\right)^{\theta / \bar{\theta}}}-
%    \frac{2^{R/ \bar{\theta}}}{\left( 1+\frac{\rho_1^\ddagger  P_s \vert h_s \vert ^2}{N}\right)^{\theta / \bar{\theta}}}
%\right)
\end{align*}
where $g_3$ is defined by \eqref{h_3} of Algorithm \ref{Algo_S_EE}.
 \label{prop:C_s}
\end{prop} 
\noindent Sub-scheme $C^{(s)}$ is specific to S-EE. The source transmits $m_r$ to the relay in Phase 1, but does not send it again in the second phase. The RD-link is strong enough to carry all information contained in $m_r$ by itself. Thus, the source saves energy by only sending new information ($m_d$) during the second phase, and does not require beamforming (which is possible if both the source and the relay transmit in the second phase). Also note that, as the RD-link is non limiting, the relay does not necessarily use all its available power to transmit $\tilde{m}_r$.

On the other hand, if the RD-link is the bottleneck of the network ($R_B \leq R_C$), the relay cannot support the transmission of $\tilde{m}_r$ alone, even by transmitting with full power. In this case, the source should repeat the message $m_r$ during Phase 2, such that beamforming gain is obtained. Then, $B^{(s)}$ is applied as follows.
\begin{prop}
Sub-scheme $B^{(s)}$ uses full decode-forward. \\
\hspace*{15pt}$\bullet$ In Phase 1, the source sends $m$, the relay decodes as $\tilde{m}$.\\
\hspace*{15pt}$\bullet$ In Phase 2, the relay sends $\tilde{m}$ and the source continues to send $m$. \\
The optimal power allocation for sub-scheme $B^{(s)}$ is 
\begin{align*}
   \rho_r^\dagger & =  \frac{1}{\bar{\theta}} \; ; \quad
    \rho_1^\dagger = \left( 2^{R / \theta }-1 \right) \frac{N}{P_s \vert h_s \vert ^2}
    \; ; \quad  
    \eta_1^\dagger  = \eta_2^\dagger = 0
     \\ %\; ; \quad
    \rho_2^\dagger  &= \frac{N}{P_s \vert h_d \vert ^2} \left( 
    \sqrt{g_3(\rho_1^\dagger )}
    - \sqrt{\frac{P_r \vert h_r \vert^2 }{ \bar{\theta}N}}
\right)^2.
\end{align*} \label{prop:B_s}
\end{prop}   %  \vspace*{-30pt}

Finally, if the source rate is very high, such that $R \geq \max \left\lbrace R^{(s)}_1,R^{(s)}_2 \right\rbrace$, partial decode-forward with beamforming is required and sub-scheme $A^{(s)}$ is applied.
\begin{prop}
In sub-scheme $A^{(s)}$, the source splits its message into two parts $m_r$ and $m_d$. \\
\hspace*{15pt}$\bullet$ In Phase 1, the source sends $m_r$ and the relay decodes as $\tilde{m}_r$. \\
\hspace*{15pt}$\bullet$ In Phase 2, the relay sends $\tilde{m}_r$ and the source sends $(m_r,m_d)$. \\
To compute the optimal power allocation set, $\rho_1^\star$ is first found numerically by solving \eqref{eq:solve_rho_1s} of Algorithm \ref{Algo_S_EE}. Next, $g_3(\rho_1^\star)$ is computed using \eqref{h_3} and $(\rho_r^\star,\rho_2^\star,\eta_1^\star,\eta_2^\star)$ are deduced from $\rho_1^\star$ and $g_3(\rho_1^\star)$  as follows:
\begin{align*}
\eta_2^\star & =  \left(\frac{2^{R/ \bar{\theta}}}{\left( 1+ \frac{\rho_1^\star P_s \vert h_s \vert ^2 }{N} \right)^{\theta / \bar{\theta}}} -1\right) \frac{N}{P_s \vert h_d \vert ^2}  
\; ; \quad
\rho_r^\star =  \frac{1}{\bar{\theta}}
\\ %
\eta_1^\star & = 0 \; ; \quad 
\rho_2^\star =  \frac{N}{P_s \vert h_d \vert ^2} \left(
%\right. \\
%& \left. 
\sqrt{ g_3(\rho_1^\star)
%\frac{2^{R/ \bar{\theta}}}{\left( 1+\frac{\rho_1^\star P_s \vert h_d \vert ^2}{N}\right)^{\theta / \bar{\theta}}}- \frac{2^{R/ \bar{\theta}}}{\left( 1+ \frac{\rho_1^\star P_s \vert h_s \vert ^2 }{N} \right)^{\theta / \bar{\theta}}}
} - \sqrt{\frac{P_r \vert h_r \vert^2}{ \bar{\theta} N}}
\right)^2
\end{align*}
\label{prop:A_s}
\end{prop}
%\vspace*{-10pt}
This sub-scheme is similar to $A^{(n)}$ and $A^{(r)}$, but with different power allocation. Even if the relay already transmits with full power, the RD-link is still too weak to support the transmission of $m_r$ alone. Thus, the source repeats the message $m_r$ during the second phase to obtain beamforming gain. Moreover, contrary to $B^{(s)}$, the SR-link is also too weak to support the transmission of the whole source message. Thus, part of the source message ($m_d$) has also to be transmitted via the direct link only.

%\vspace*{5pt}
\textbf{{Existence of the sub-schemes: }}
Considering the optimized scheme as function of the source rate, S-EE is composed of either $(D^{(s)},B^{(s)},A^{(s)})$ or $(D^{(s)},C^{(s)},A^{(s)})$, in that order, depending on which link is the bottleneck of the network.
%Figure \ref{Fig:Example_subschemes_source} illustrates the sequence $(D^{(s)},B^{(s)},A^{(s)})$ and plots the total energy consumption during the second phase as a function of the source rate.
Depending on the power constraints and channel gains, S-EE can also be composed of fewer than three sub-schemes and outage can occur with any sub-scheme $D^{(s)}$, $C^{(s)}$, $B^{(s)}$ or $A^{(s)}$.
Specifically, when the gains of the SD- and SR-links are close, sub-scheme $D^{(s)}$ (two-hop relaying) may not be applied at all. Indeed, in this case and even for low source rates, two-hop relaying under-uses the direct link to transmit information and is thus sub-optimal.
Furthermore, given the power constraints, outage can occur before the sequences $(D^{(s)},B^{(s)},A^{(s)})$ and $(D^{(s)},C^{(s)},A^{(s)})$ are completed, such that $A^{(s)}$, $B^{(s)}$ or $C^{(s)}$ may not be applied. We next describe the condition of existence of these sub-schemes.

First, consider the sequence $(D^{(s)},C^{(s)},A^{(s)})$, i.e. two-hop relaying, full decode-forward without beamforming and partial decode-forward with beamforming. When the SR-link is the limiting one, S-EE is composed of only $(D^{(s)},C^{(s)})$ for most channel realizations. Partial decode-forward with beamforming ($A^{(s)}$) is only applied if the relay power $P_r$ is very low compared to the source power $P_s$.

Second, we consider the sequence $(D^{(s)},B^{(s)},A^{(s)})$, i.e. two-hop relaying, full decode-forward with beamforming and partial decode-forward with beamforming. When the network is severely limited by both the SD- and RD-links, S-EE is composed of only $(D^{(s)},B^{(s)})$. Recall that in $B^{(s)}$, the relay constraint is already met. Thus, if the direct link is weak given the source power constraint, the scheme rapidly goes into outage and cannot achieve high rates.

Finally, S-EE can also be reduced to only two-hop relaying ($D^{(s)}$) when the direct link is almost null.

%%%%%%%%%%%%%%%%%%%%%%%%%%%%%%%%%%%%%%%%%%%%%%%%%%%%%%%%%%
\subsection{Maximum rates achieved by the three optimized schemes}
\label{prop:scheme_comp}

We now analyze the maximum rate achievable for the three optimized schemes. Formal proof can be found in Appendix \ref{appendix:Maximal_achievable_source_rate}. 
First, in N-EE (which minimizes the network consumption), the source power constraint is always met with equality before the relay power constraint. Therefore, at the expense of consuming more total energy, S-EE and R-EE (which respectively minimize the source and the relay consumption) allow the source to reach higher rates than the maximal rate achieved by scheme N-EE, denoted as $R^{(n)}_{\max}$.

Second, analyzing the outage S-EE sub-schemes gives a comprehensive description of the overall maximum achievable rate $R_{\max}$. Outage is declared when the power constraints cannot be met given the source rate, which can occur with any sub-scheme, as explained in Section \ref{sec:S-EE}.
For sub-schemes $A^{(s)}$ and $B^{(s)}$, the relay power constraint is already met and S-EE goes in outage as soon as the source power constraint is met. Therefore, in this case, at $R=R_{\max}$, both source and relay power constraints are met.
On the contrary, when sub-schemes $C^{(s)}$ and $D^{(s)}$ are applied, the relay power constraint is not limiting. Thus, S-EE goes in outage only if the source power constraint is met. In this case, the maximum achievable rate $R_{\max}$ can be derived in closed-form. The values of $R_{\max}$ are described in Table \ref{table:R_max}. Figure \ref{Fig:Power_consumption} gives an example of $R_{\max}$.

Last, given power constraints and channel gains, S-EE and R-EE schemes reach the same maximum achievable rate $R_{\max}$ in Table \ref{table:R_max} and have the same optimal power allocation at this specific rate.

\newcounter{algorithmcounter}
\setcounter{algorithmcounter}{\value{table}}
\addtocounter{algorithmcounter}{1}

\newcounter{tableau}
\setcounter{tableau}{1}

\renewcommand\thetable{\Roman{tableau}}
\renewcommand{\tablename}{\textsc{Table}} 
\begin{table*} \centering 
\begin{tabular}{c|c|c}
Channel conditions & Outage with sub-scheme & $R_{\max}$ \\ \hline 
$\left.\begin{array}{r}
\text{Weak SR-link and low } P_r \\
\text{or Weak RD-link} 
\end{array}\right\rbrace$
 & $A^{(s)}$ & 
$I_1$ in \eqref{eq:gauss_const}, with $\eta_1$, $\rho_1$, $\eta_2$ and $\rho_2$ of $A^{(s)}$ and $\rho_r = \frac{1}{\bar{\theta}}$ \\
%or Weak RD-link & &
%\\ 
Weak SD- and RD-links & $B^{(s)}$ &
$I_1$ in \eqref{eq:gauss_const},  with $\eta_1$, $\rho_1$, $\eta_2$ and $\rho_2$ of $B^{(s)}$ and $\rho_r = \frac{1}{\bar{\theta}}$\\ 
Weak SR-link & $C^{(s)}$ &
$\log_2 \left(\theta + \frac{\vert h_s \vert ^2}{\vert h_d \vert ^2} \bar{\theta} +  \frac{P_s \vert h_s \vert ^2}{N} \right) + \bar{\theta} \log_2 \left( \frac{\vert h_d \vert ^2}{\vert h_s \vert ^2} \right) $ \\
SD-link close to 0 & $D^{(s)}$ & $\theta \log_2 \left(1+ \frac{P_s \vert h_s \vert^2}{\theta N} \right)$ \\
\end{tabular}
\caption{Maximum achievable rate $R_{\max}$ \vspace*{-10pt}}  %\vspace*{-20pt} 
\label{table:R_max}
\end{table*}
\addtocounter{tableau}{1}
\renewcommand\thetable{\Roman{algorithmcounter}}
\renewcommand{\tablename}{\textsc{Algorithm}}

\section{A generalized scheme for energy efficiency (G-EE)}
\label{sec:G-EE}

So far, we have proposed three optimized schemes N-EE, R-EE and S-EE with power allocations that respectively minimize the network, the relay and the source energy consumption. We now propose a generalized scheme that combines these optimal power allocations in a smooth way. % \vspace*{-8pt}

\subsection{Implementing a generalized scheme (G-EE)}

As shown in Section \ref{prop:scheme_comp}, N-EE sub-schemes are insufficient to describe the whole range of achievable rates and both R-EE and S-EE can increase this range.

Considering power consumption as a function of the source rate, the resource allocation that minimizes the network consumption (N-EE) is extended beyond its maximal achievable rate $R^{(n)}_{\max}$ by the allocation that minimizes the relay consumption (R-EE) in a continuous and differentiable manner (see Appendix \ref{appendix:diff} for the detailed proof).
Therefore, we propose a generalized scheme that combines these allocations and allows higher achievable rates.
In this generalized scheme, denoted as G-EE, the message splitting and power allocation for minimizing the total energy consumption are applied as long as feasible, given the rate and power constraints. In particular, $B^{(n)}$ (full decode-forward) is used for source rates lower than $R^{(n)}$, and $A^{(n)}$ (partial decode-forward) is used for rates above $R^{(n)}$ and up to $R^{(n)}_{\max}$, which is computed by $I_1$ in \eqref{eq:gauss_const} with equalized source power constraint \eqref{eq:power_const}.
Once the source rate is infeasible for N-EE, the message splitting and power allocation for minimizing the relay energy consumption alone is applied, by using $A^{(r)}$ (partial decode-forward).

\begin{table}
\centering \fbox{
\begin{minipage}{0.9\columnwidth}
\begin{algorithmic}
	\IF{$R\leq R^{(n)}$}
	\STATE Apply $B^{(n)}$ (full DF for network energy-efficiency)
	\ELSIF{$R^{(n)}\leq R \leq R^{(n)}_{\max}$}
		\STATE Apply $A^{(n)}$ (partial DF for network energy-efficiency)
	\ELSIF{$R^{(n)}_{\max}\leq R \leq R_{\max}$}
		\STATE Apply $A^{(r)}$ (partial DF for relay energy-efficiency)
		\ELSE
		\STATE Declare outage
	\ENDIF
\end{algorithmic}
\end{minipage}
}
\caption{Generalized optimal scheme for energy efficiency, G-EE \vspace*{-15pt}} %\vspace*{-20pt}
\label{Algo_G_EE}
\end{table}

\subsection{Comparison between G-EE and rate-optimal schemes}

Considering rates as a function of energy, G-EE is also rate-optimal for the consumed energy in the half-duplex relay channel with decode-forward.
\begin{proof}
If a source rate achieved with G-EE is not rate-optimal for this energy, then there exists another set of power allocation that consumes the same energy but achieves a better rate $R_m$. For this higher rate $R_m$, G-EE consumes more energy that the maximum-rate scheme and is thus suboptimal for energy, which is contradictory. Thus, for all achievable rates, G-EE is also rate-optimal. Also G-EE achieves all the rates that are achievable by the maximum-rate scheme (see below).
\end{proof} 

Considering energy as a function of rates and given fixed individual power constraints, the maximum source rate $R_{\max}$ achieved with energy minimization G-EE is equal to the maximum rate achieved by rate maximization.
\begin{proof}
Note that both G-EE and the maximum-rate scheme are optimization problems with four constraints (two rate and two power constraints). The two rate constraints are always met with equality in both problems and, with source rate up to $R_{\max}$, G-EE is equivalent to the rate-optimal scheme. When $R \rightarrow R_{\max}$, the source power constraint is also met with equality (G-EE applies R-EE). Hence, in the interior neighbourhood of $R_{\max}$, each optimization problem has just one degree of freedom left. When $R \geq R_{\max}$, an outage occurs: either the desired source rate becomes infeasible, or the relay power constraint is also met with equality. Thus, above $R_{\max}$, there is no more degree of freedom for neither problem and $R_{\max}$ is the maximal rate that is achieved for both G-EE and maximum-rate schemes.
\end{proof} 

Finally, both previous points imply that only the combined scheme G-EE is equivalent to the maximum-rate scheme.
Figure \ref{Fig:Power_consumption} illustrates this equivalence. 
As a reference, we consider the rate-optimal scheme proposed in \cite{Host-Madsen,Host-MadsenJournal}. This scheme is similar to $A^{(n)}$ in coding: the source performs message splitting, sending one part during phase 1, and both during phase 2. Then, the power allocation is numerically optimized to maximize the achievable rate.
For the comparison, we compute the maximum capacity achievable with the rate-optimal scheme based on the energy consumed using G-EE (plot with reversed axis).

\subsection{Discussion}

Even though both analysis and simulation show that the generalized scheme G-EE and the maximum-rate scheme proposed in \cite{Host-Madsen,Host-MadsenJournal} lead to the same set of power allocation, we argue that the proposed scheme G-EE brings two significant advantages.

First, existing maximum-rate schemes proposed in the literature require numerical search of the optimal solution for each channel realization. On the contrary, the G-EE scheme, which is both rate- and energy-optimal, has the optimal power allocation in closed-form.

Second, this energy efficiency approach reveals more details about sub-schemes and power allocations than the maximum-rate approach. We show that minimizing the network energy consumption is not equivalent to maximizing the network capacity as often believed, since the whole range of source rates achievable with decode-forward cannot be covered. Furthermore, we highlight that several regimes of coding technique and power allocation exist (partial or full decode-forward, with or without beamforming) and each of them corresponds to a specific energy optimization problem (N-EE or R-EE).

%%%%%%%%%%%%%%%%%%%%%%%%%%%%%%%%%%%%%%%%%%%%%%%%%%%%%%%%%%

\section{Simulation results}
\label{sec:simulations}

\subsection{Description of the simulation environment} \label{sec:environment}

We simulate the performance of the proposed schemes in a realistic environment and consider both pathloss and shadowing. Shadowing is the most severe random attenuation factor encountered in urban cellular environment \cite{livre_LTE}. It refers to mid-term channel variations which are due to the multipath propagation generated by obstructions between the transmitter and receiver. Since blocking objects do not vary rapidly in geographical position, size or dielectric properties, shadowing remains slow and predictable. Therefore, in our case, we can reasonably assume that channel gains are constant over the two transmission phases and that nodes can track those variations at each instant.

\begin{figure}
\centering \includegraphics[width = 0.5\columnwidth]{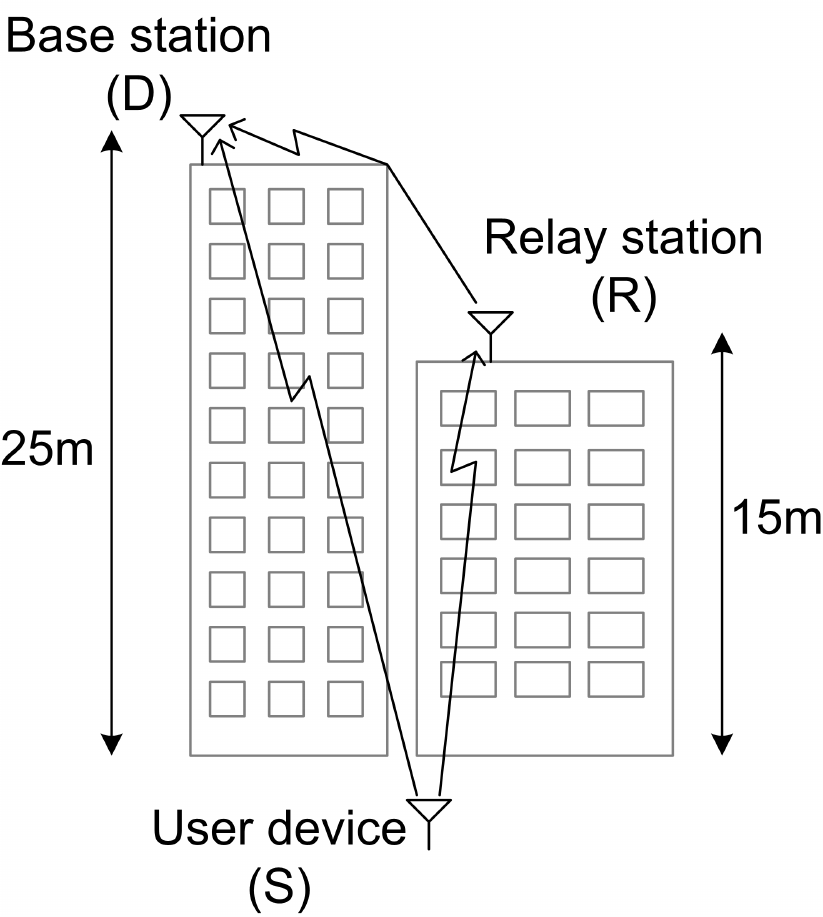} 
\caption{Communication in a realistic environment}
\label{real_situation}
\end{figure}

We consider the situation depicted in Figure \ref{real_situation}. In a dense urban environment with high surrounding buildings, a mobile user (source S) sends data to a base station (destination D) with the help of a fixed relay station (relay R), such as in LTE or Wimax. The mobile user is located outdoors, at street level. The relay station is below the top of surrounding building and the base station is far above the rooftop. This channel model corresponds to the Stationary Feeder scenario described in the European WINNER project \cite{Winner}. Pathloss and shadowing effects are given by (4.23) of \cite{Winner}:
%\vspace*{5pt}
\begin{align*}
\text{PL} = A \log_{10}(d\text{[m]}) + B + C \log_{10}\left( \frac{f_c \text{[GHz]}}{5.0} \right) + \chi
\end{align*}
%\vspace{5pt}
\noindent where $d$ is the distance between the transmitter and receiver and $f_c$ is the frequency carrier (within the range 2-6GHz). $\chi$ refers to shadowing effect, it is modelled as a zero-mean log-normal random process, with standard deviation $\sigma^2$. The parameters $A$, $B$, $C$ and $\sigma^2$ depend on the global location of the transmitters and receivers (street level, rooftop...).
Considering the relative positions of nodes and using notations of \cite{Winner}, the SR-link is modelled by scenario B1, the SD-link by scenario C2 and the RD-link by scenario B5f. Table \ref{pathloss} gives the distances transmitter-receiver, as well as the pathloss and shadowing parameters for each link.
For simulations, we take $f_c=3.5$GHz, $P_s=100$mW, $P_r=200$mW, which refers to a Wimax situation. We also consider $\theta = 0.3$ and $N=10^{-10}$W.

\renewcommand\thetable{\Roman{tableau}}
\renewcommand{\tablename}{\textsc{Table}} 

\begin{table}\normalsize
%\begin{minipage}{0.45\textwidth}
      \centering
       \begin{tabular}{|c|c|c|c|c|c|}
		\hline
		Link & d[m] & $A$ & $B$ & $C$ & $\sigma^2$ \\ \hline
		SR & 70 & 22.7 & 41.0 & 20 & 3 \\ \hline
		SD & 100 & 26 & 39 & 20 & 4 \\ \hline
		RD & 30 & 23.5 & 57.5 & 23 & 8 \\ \hline
		\end{tabular} 
		\caption{Pathloss and shadowing parameters}
				\label{pathloss}
\addtocounter{tableau}{1}
% \vspace*{-20pt}
\end{table}
\renewcommand{\tablename}{\textsc{Algorithm}}

\subsection{Reference Schemes}
\label{reference}

As reference schemes, we consider direct transmission, two-hop transmission and the maximum-rate scheme in \cite{Host-Madsen,Host-MadsenJournal}. 
Direct transmission is one-phase communications. The source sends the message $m$ with no splitting ($m_d = m$ and $m_r = 0$) and minimum power
\begin{align*}
\eta_D P_s = \frac{\left( 2^{R} -1 \right) N}{\vert h_d \vert ^2} \; .
\end{align*}
%
%\subsubsection{Two-Hop Routing}
In the classical two-hop routing, the source first sends the whole message to the relay, again without splitting, and with a minimum power $ \rho^{S}_{2H} P_s$. If the relay can decode the message, it forwards to the destination with a power of $ \rho^{R}_{2H} P_r$, where
\begin{align*}
\rho^{S}_{2H} P_s = \frac{\left( 2^{R/\theta} -1 \right) N}{\vert h_s \vert ^2}  \quad ; \quad
\rho^{R}_{2H} P_r= \frac{\left( 2^{R/\bar{\theta}} -1 \right) N}{\vert h_r \vert ^2} \; .
\end{align*}
The destination then decodes using only the signal received from the relay (not both signals received from source and relay as in the two-hop sub-scheme $D^{(s)}$).

\begin{figure}
\begin{center}
\includegraphics[width=0.65\columnwidth]{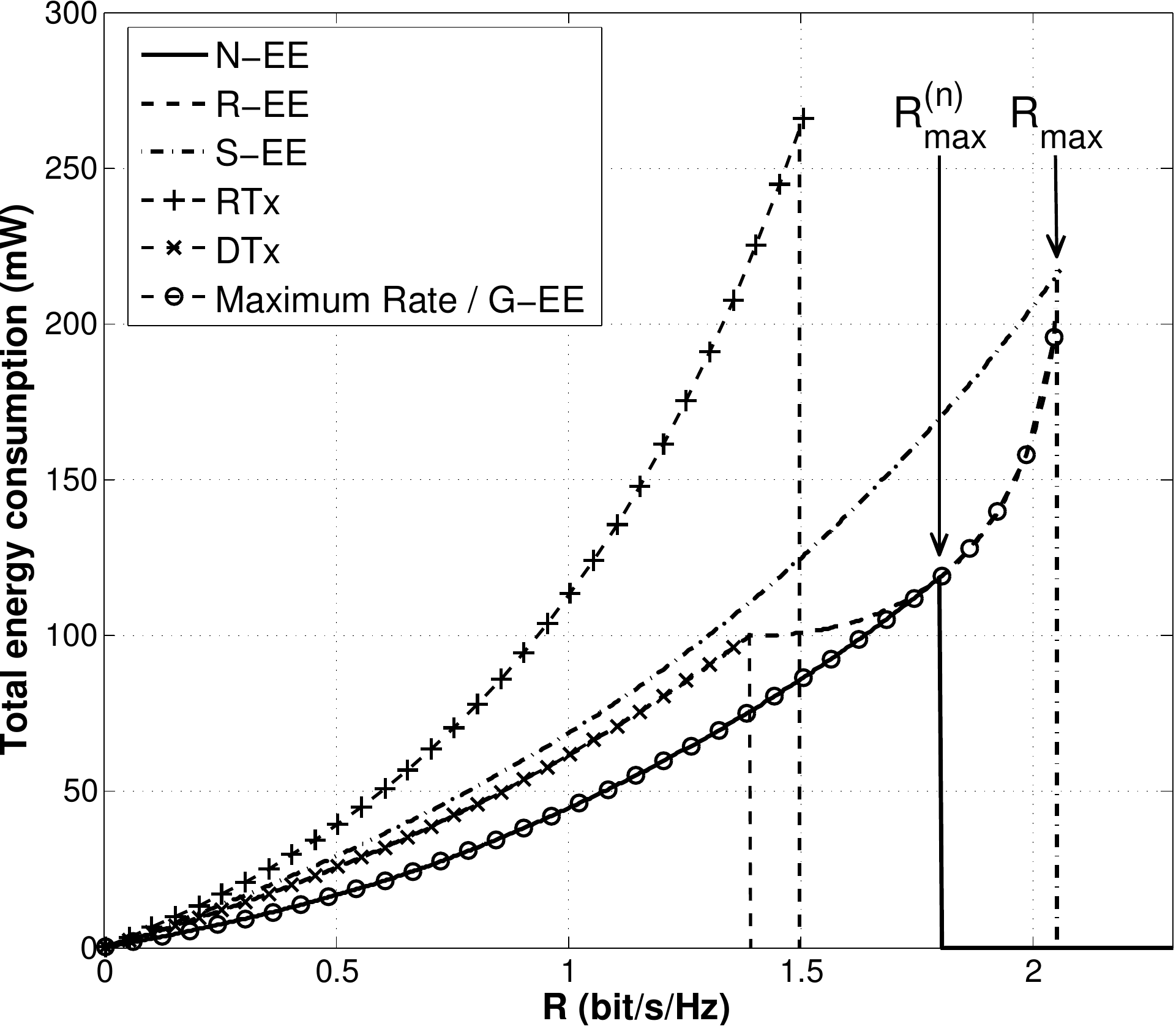}
\end{center}
\caption{Total power consumption as function of the rate during both phases.}
\label{Fig:Power_consumption}
\end{figure}

The maximum-rate scheme in \cite{Host-Madsen,Host-MadsenJournal} uses partial decode-forward with power allocation optimal for rate. Allocation is computed numerically solving a convex optimization problem. Whereas classical performance simulations plot capacity as a function of energy, we reverse the axes to plot energy as a function of rate and to allow comparison.
We then analyze the performance of the proposed schemes in both non-fading and fading environments.

%\vspace*{-10pt}
\subsection{Performance in non fading environment}

We first consider a non fading environment, with only pathloss ($\chi$ is set to zero). This can also be viewed as instantaneous performance.
%
%\subsubsection{Power Consumption and Source Rate}

\begin{figure}
\begin{center}
\includegraphics[width=0.65\columnwidth]{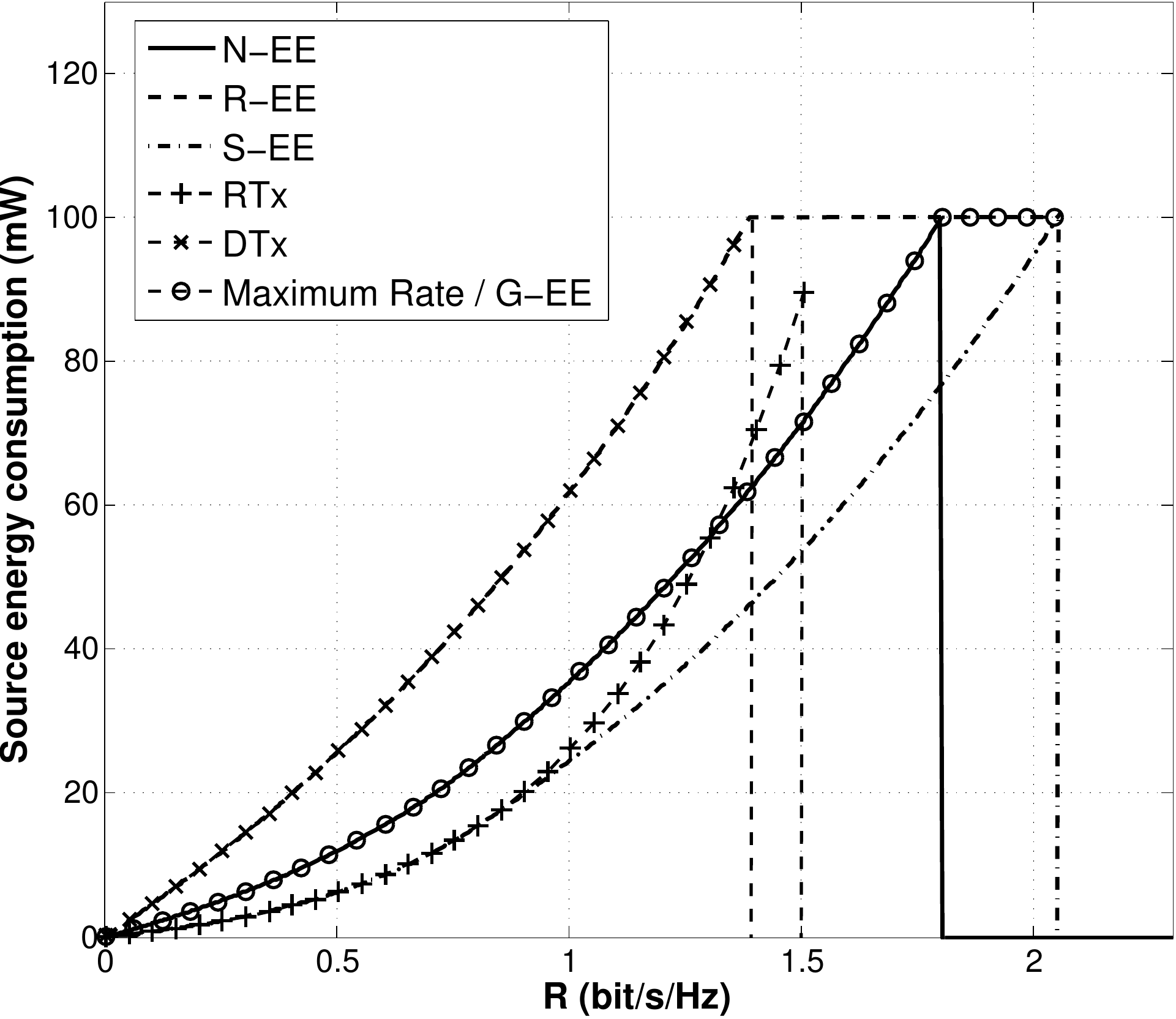}
\includegraphics[width=0.65\columnwidth]{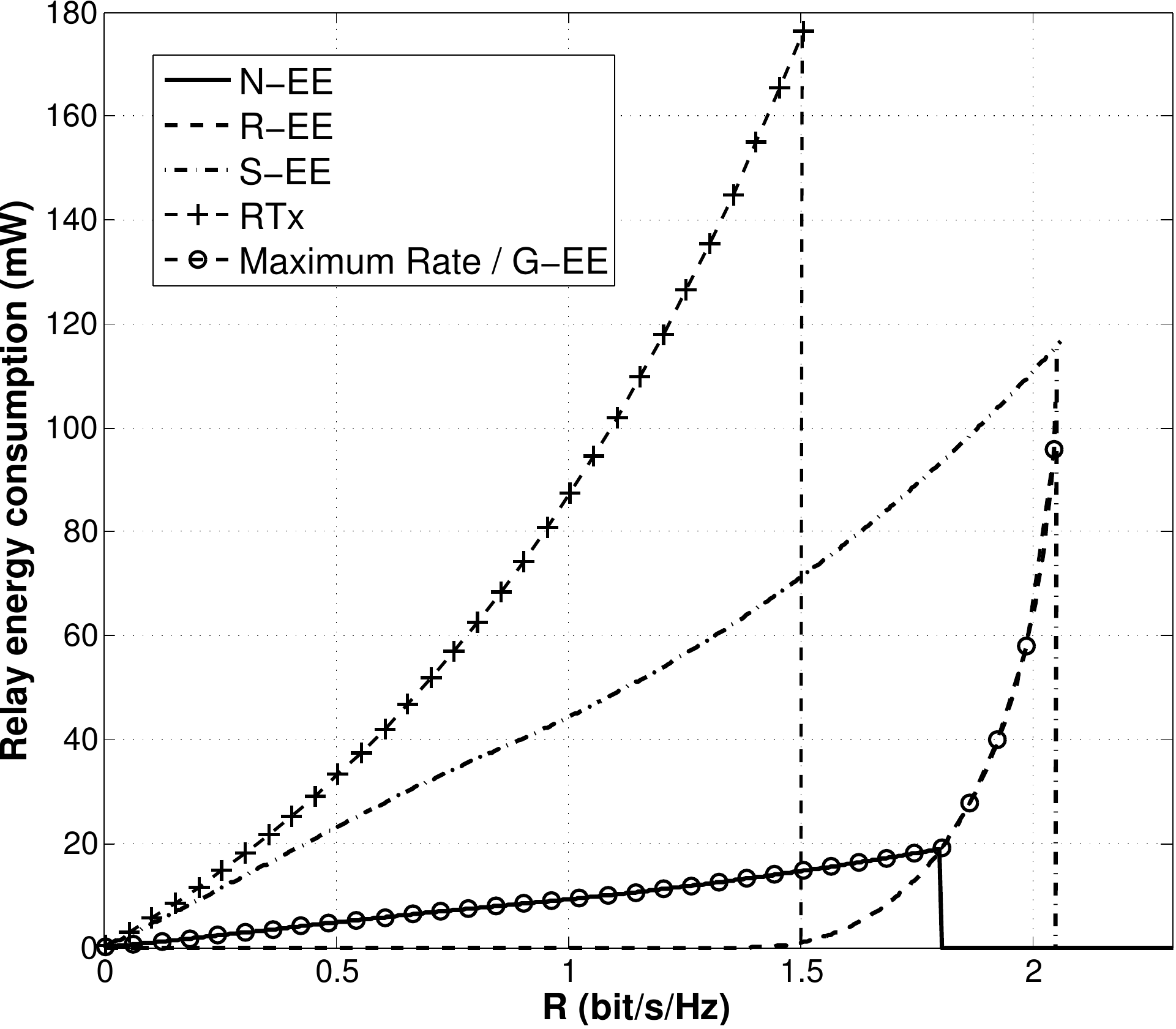} 
\end{center}
\caption{Power consumption per node: source (above) and relay (below)}
\label{Fig:Power_consumption_per_node}
\end{figure}

Figures \ref{Fig:Power_consumption} and \ref{Fig:Power_consumption_per_node} respectively show the total power and the per-node power needed to maintain a range of desired source rates. When a source rate is not achievable, an outage occurs, which is depicted by a cut-off in the curve.
These figures illustrate the analysis of Section \ref{prop:scheme_comp}. First, higher rates are achieved by optimizing only one node consumption (either the source as in S-EE, or the relay as in R-EE), rather than the whole network consumption (N-EE).
Moreover, both schemes S-EE and R-EE achieve the same maximum source rate.
Second, the combined scheme G-EE is also rate-optimal and significantly outperforms both direct and two-hop routing for every achievable source rate. For example, at a rate of 1.35 normalized b/s/Hz, 1.2dB of energy gain is obtained over direct transmissions and 4.5dB over two-hop routing.
The gain comes mainly from lessening the relay power consumption. Note that at $R=R_{\max}$, the relay power constraint is not necessarily met with equality if rates beyond $R_{\max}$ are infeasible due to rate constraints. Similarly, R-EE and S-EE provide significant power gain at the relay and the source respectively. Note that maximum-rate scheme is neither source nor relay energy-efficient as it consumes more power at each respective node than the optimal scheme for that node.

\begin{figure}
\begin{center}
\includegraphics[width=0.65\columnwidth]{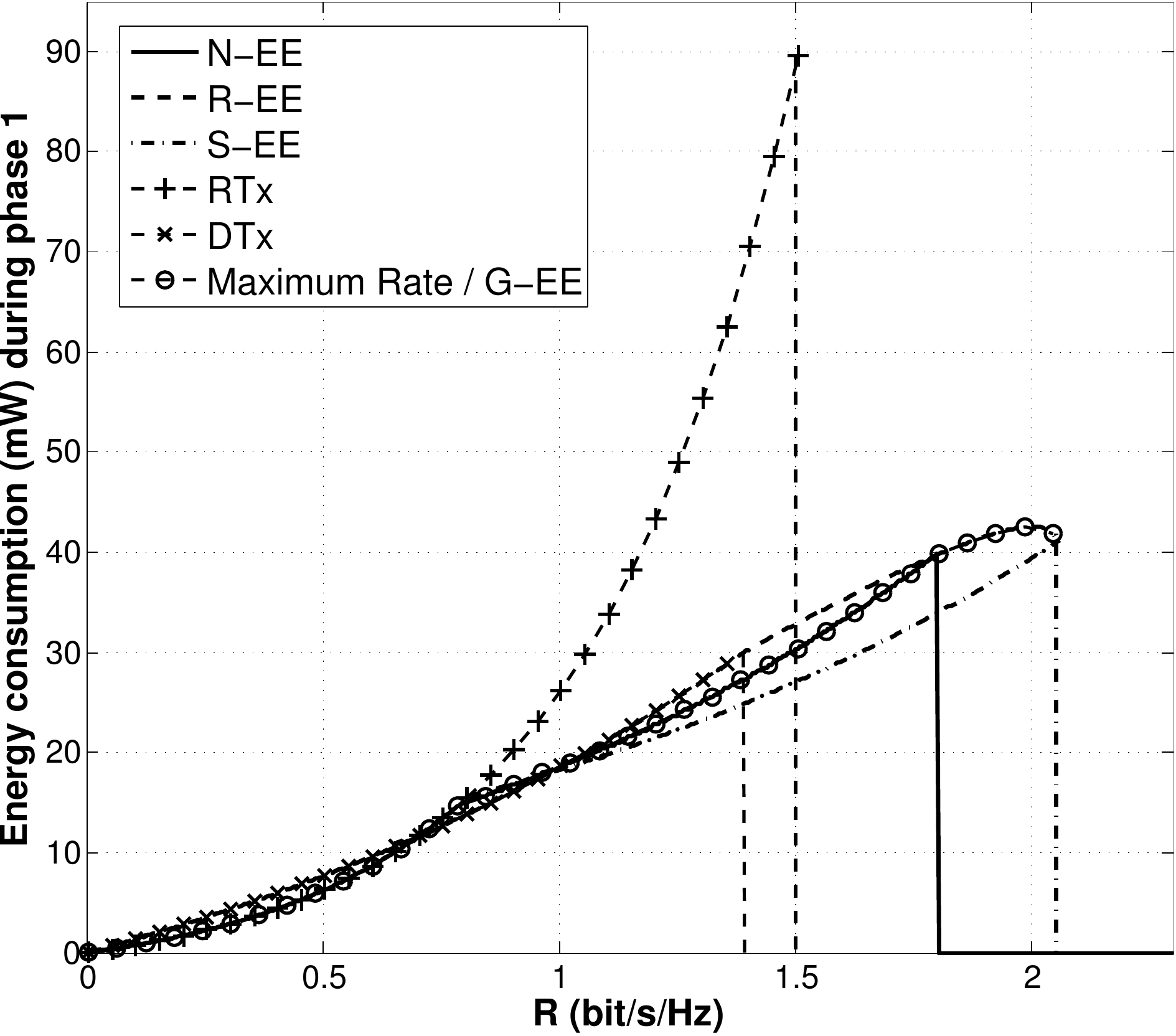} 
\includegraphics[width=0.65\columnwidth]{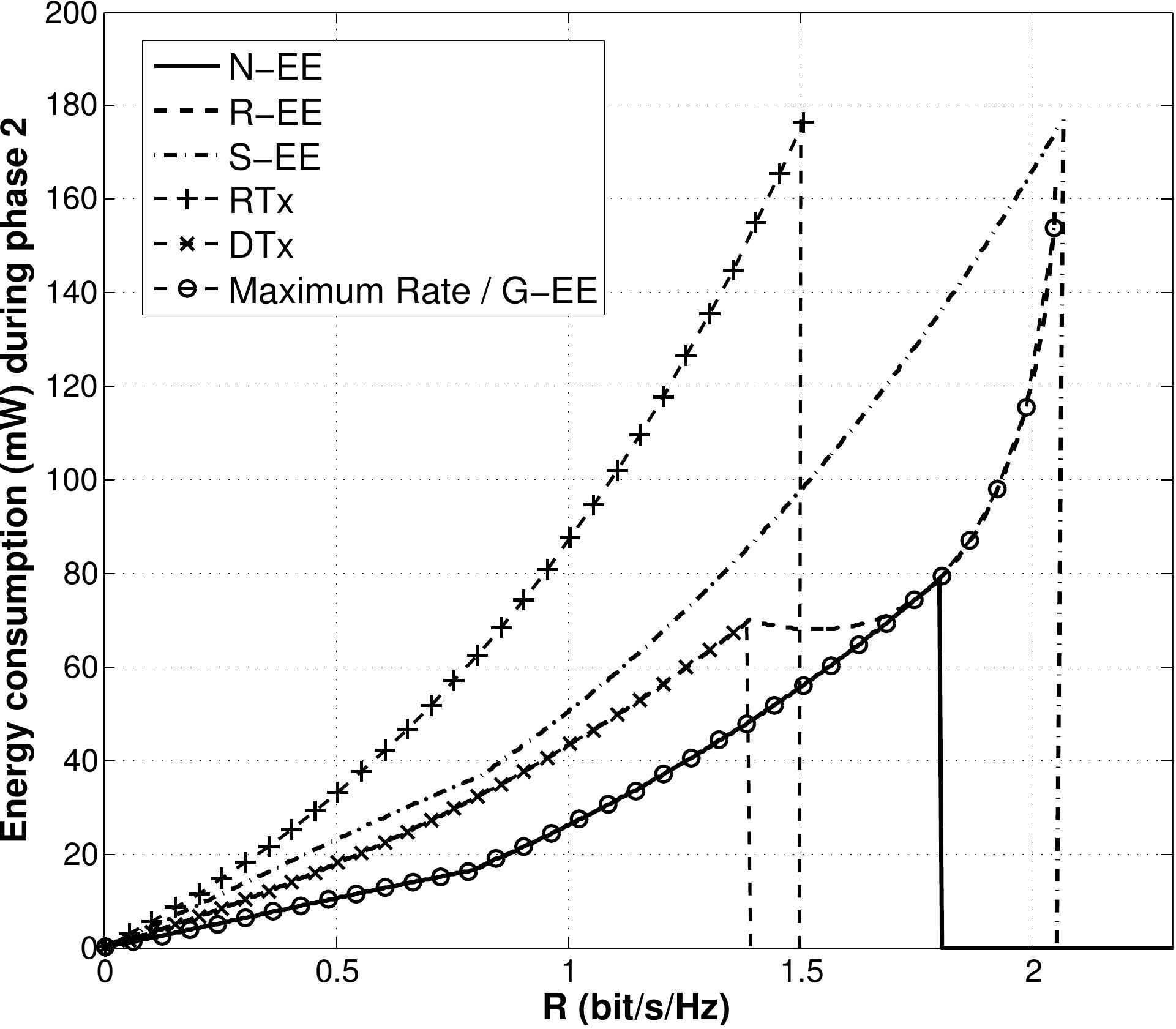} 
\end{center}
\caption{Power consumption per phase: phase 1 (above) and phase 2 (below)}
\label{Fig:Power_consumption_per_phase}
\end{figure}

\begin{figure}
\begin{center}
\includegraphics[width=0.65\columnwidth]{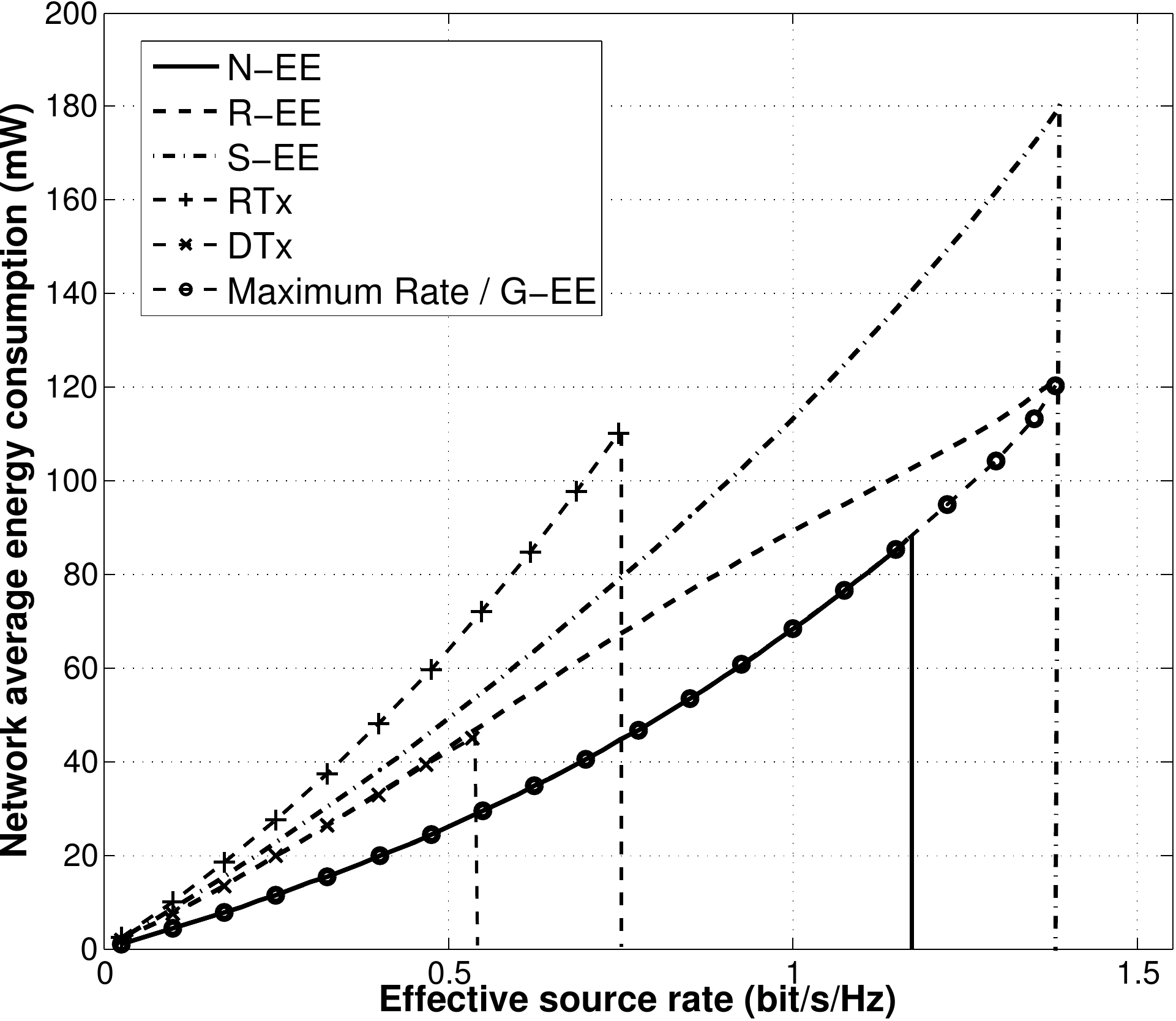}
\end{center}
\caption{Total average energy gain as a function of the source rate}
\label{Fig:fading_total}
\end{figure}

Two-hop routing generally suffers from very high instantaneous power consumption. Since the whole message is sent twice, one by the source and one by the relay, within the same time slot, both transmitters have to increase their transmit power to sustain the desired rate. We plot in Figure \ref{Fig:Power_consumption_per_phase} the power consumed at each phase for the proposed and reference schemes.
We see that the proposed schemes allocate power in a smooth manner and reduce transmit power peaks. Contrary to routing, energy is more evenly spread over both phases and over both direct and relaying paths. Moreover, the consumption per phase as function of the source rate is similar to the direct transmission consumption. This can help managing interference in networks which allow both direct and relayed transmissions.

\subsection{Performance in fading environment}

We consider in this section both pathloss and shadowing. Due to random shadowing, there exists a non-zero probability that the channel gains are too weak to perform transmission. We plot in Figure \ref{Fig:fading_total} the average total network energy consumption. The cut-off in the curve depicts the maximum rate above which the outage probability is greater than 0.05. Simulations show that the proposed schemes significantly reduce the average energy consumed, as well as increase the maximal achievable source rate, given the outage requirement.

\begin{figure}
\begin{center}
\includegraphics[width=0.75\columnwidth]{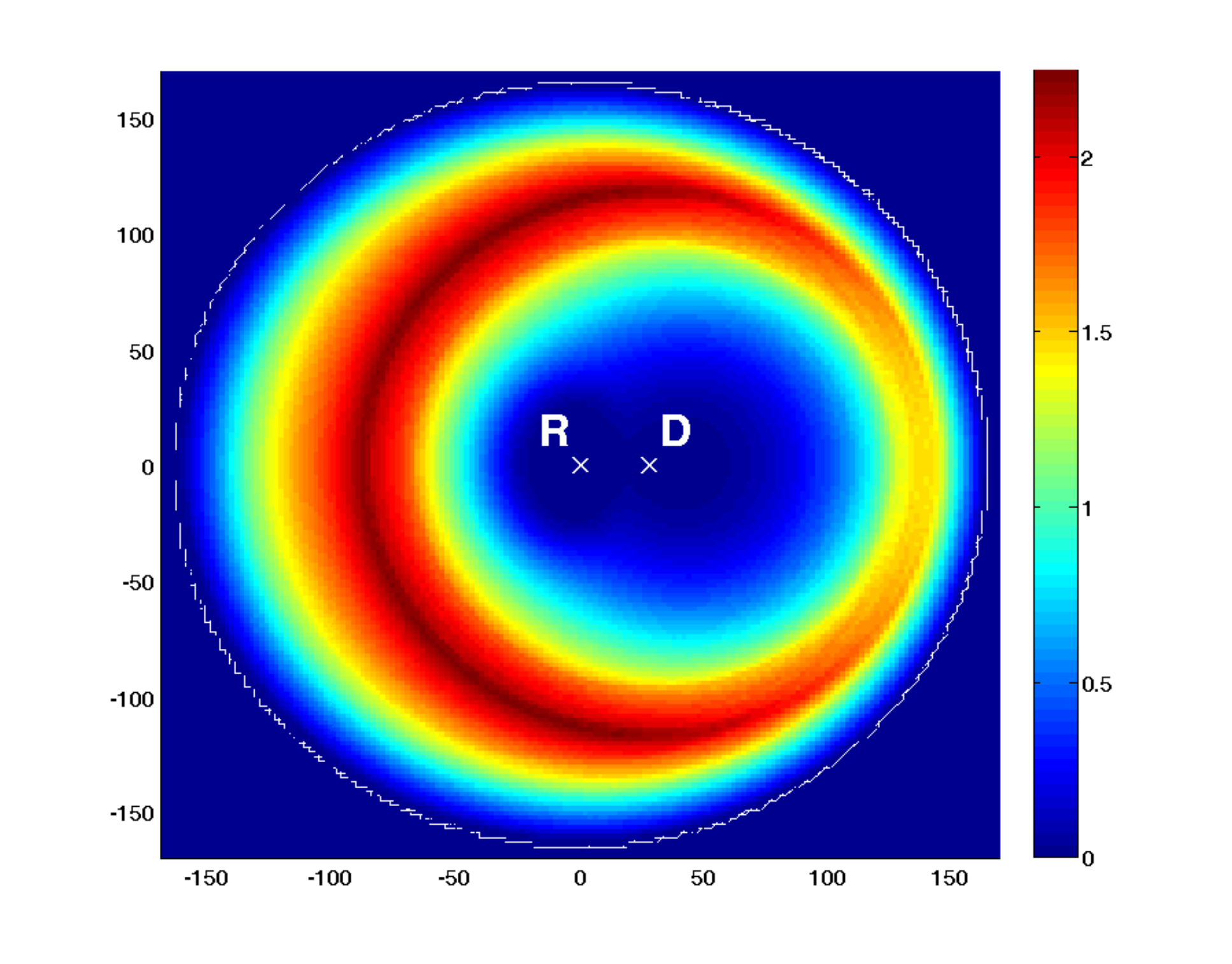}
\end{center}
\vspace*{-20pt}\caption{Total average energy gain (dB) as function of the user position}
\label{Fig:Network_gain_position_fading}
\end{figure}

\begin{figure}
\begin{center}
\includegraphics[width=0.75\columnwidth]{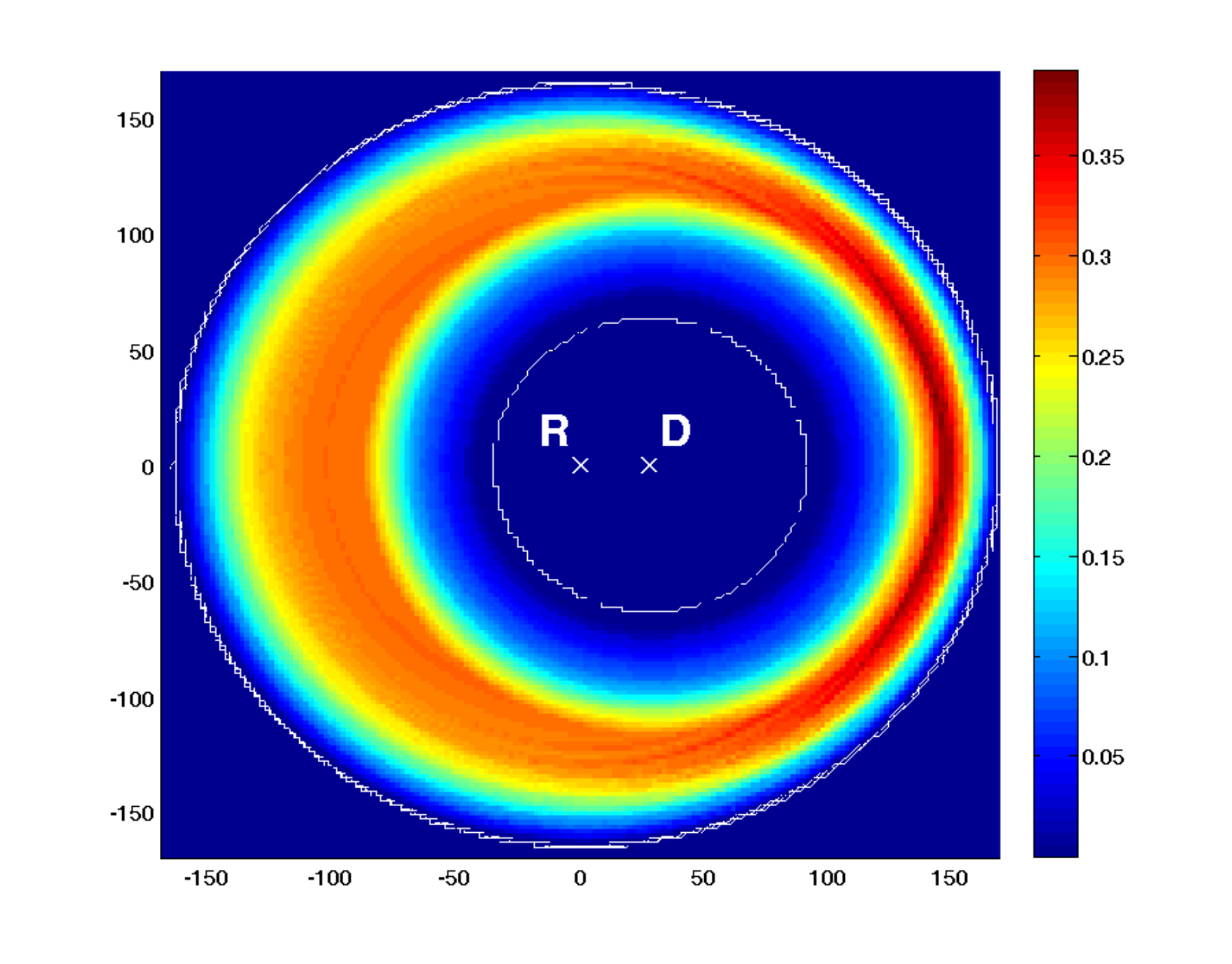}
\end{center}
\vspace*{-20pt}\caption{Outage reduction (\%) as a function of the mobile user position}
\label{Fig:Network_outage_position_fading}
\end{figure}

Finally, we analyze the utilization region of the proposed schemes and evaluate the energy gain as a function of the mobile user location, similarly to \cite{art2-7}.
We consider the environment and heights of relay and base station as described in Section \ref{sec:environment}. The mobile user is free to locate anywhere at street level. The relay station is located at the origin $(0,0)$ and the destination at $(20\sqrt{2},0)$ so that $d_{\text{RD}}=30$m.
As reference, we consider a combined scheme which uses either direct transmissions (DTx) or two-hop routing (RTx), depending on the total consumed energy. %Note that this reference scheme is applied in most standards using relays, such as LTE or Wimax.
For simulations, equal time division is assumed, with a source rate of 1bit/s/Hz.
Figures \ref{Fig:Network_gain_position_fading} and \ref{Fig:Network_outage_position_fading} respectively depict the power gain in dB realized by the proposed generalized scheme G-EE over the reference combined scheme and the outage gain (computed as the difference between outage probabilities of reference and G-EE). 
Simulation shows that G-EE  provides up to 2.2dB of energy gain, as well as decreases the outage probability by up to 37.5\%. G-EE is particularly beneficial at cell-edge, for both energy and outage.

%%%%%%%%%%%%%%%%%%%%%%%%%%%%%%%%%%%%%%%%%%%%%%%%%%%%%%%%%%
\section{Conlusion}
\label{sec:conclusion}

We explore the issue of energy efficiency in relay channels and design a half-duplex coding scheme with optimal resource allocations that maintain a desired source rate and minimize the energy consumption. We consider the network, the relay and the source consumption separately and then combine them into a generalized set of power allocation G-EE which is energy optimal for the relay channel.

We show that minimizing the network energy consumption is not equivalent to maximizing the network capacity as often believed, since the whole range of source rates achievable with Decode-Forward cannot be covered by individual schemes. This equivalence only holds for the proposed generalized scheme G-EE.

Furthermore, this energy efficiency approach remains meaningful since it allows a comprehensive description of the optimal coding (full or partial decode-forward, with or without beamforming) as well as closed-form solutions of the optimal power allocation, which remains implicit in maximum-rate schemes.  Reversely, we highlight that rate-optimal scheme is not energy-efficient for either node. 

The generalized scheme G-EE gives a practical algorithm for message coding and associated power allocation to reach energy efficiency and rate optimality for each desired source rate and channel realization. Applied in cellular model, G-EE is particularly beneficial at cell-edge where up to 2.2dB energy gain can be obtained, as well as a reduction of 37.5\% of the outage probability.
%\vspace*{-5pt}

%%%%%%%%%%%%%%%%%%%%%%%%%%%%%%%%%%%%%%%%%%%%%%%%%%%%%%%%%%
\appendices

%%%%%%%%%%%%%%%%%%%%%%%%%%%%%%%%%%%%%%%%%%%%%%%%%%%%%%%%%%%%%%%%%%%%%%%%%%%%%%%
\begin{figure*}[!t]
% ensure that we have normalsize text
\normalsize
% Store the current equation number.
\setcounter{MYtempeqncnt}{\value{equation}}
% Set the equation number to one less than the one
% desired for the first equation here.
% The value here will have to changed if equations
% are added or removed prior to the place these
% equations are referenced in the main text.
\setcounter{equation}{13}
\small{
\begin{align}
\hspace*{-10pt} \frac{\partial \mathcal{L}}{\partial \eta_1} 
	& =	\omega_s \theta P_s
	- \frac{\lambda_1 \theta  \frac{P_s \vert h_d \vert ^2 }{N}}{\left( 1+ \frac{\left( \eta_1+\rho_1 \right)P_s \vert h_d \vert ^2 }{N}\right)}
	- \frac{\lambda_2 \theta P_s }{  N} \left(\frac{ \vert h_s \vert ^2 }{\left( 1+ \frac{\left( \eta_1+\rho_1 \right)P_s \vert h_s \vert ^2 }{N}\right)}
	-  \frac{\vert h_s \vert ^2 }{\left( 1+ \frac{\eta_1 P_s \vert h_s \vert ^2 }{N}\right)}
%	\right. \nonumber \\ & \left. 
%	\quad \quad \quad \quad \quad \quad \quad \quad \quad \quad \quad \quad \quad \quad \quad \quad \quad \quad \quad \quad \quad 
%	 \quad  \quad \quad \quad \quad \quad
	+ \frac{\vert h_d \vert ^2}{\left( 1+ \frac{\eta_1 P_s \vert h_d \vert ^2 }{N}\right)} \right) 
=0 \label{eq:lag_eta1}
\\
\hspace*{-10pt}\frac{\partial \mathcal{L}}{\partial \rho_1} 
	& = \omega_s \theta P_s
	- \lambda_1 \frac{\theta \frac{P_s \vert h_d \vert ^2 }{N}}{ \left( 1+ \frac{\left( \eta_1+\rho_1 \right)P_s \vert h_d \vert ^2 }{N}\right)}
	-  \lambda_2 \frac{\theta P_s }{  N} \frac{ \vert h_s \vert ^2 }{\left( 1+ \frac{\left( \eta_1+\rho_1 \right)P_s \vert h_s \vert ^2 }{N}\right)}
=0 \label{eq:lag_rho1}
\\
\hspace*{-10pt}\frac{\partial \mathcal{L}}{\partial \eta_2} 
	& = \omega_s \bar{\theta} P_s
	- \lambda_1 \bar{\theta} \frac{\frac{P_s \vert h_d \vert ^2}{N}  }
		{  \left( 1+ \frac{\left( \eta_2+\rho_2 \right)P_s \vert h_d \vert ^2 + \vert h_r \vert ^2 \rho_r P_r  + 2 \sqrt{P_s \vert h_d \vert ^2 P_r  \vert h_r \vert ^2 \rho_2 \rho_r}}{N}\right)}
	- \lambda_2 \bar{\theta} \frac{\frac{P_s \vert h_d \vert ^2}{N}  }{\left( 1+ \frac{\eta_2 P_s \vert h_d \vert ^2 }{N}\right)}
=0 \label{eq:lag_eta2}
\\
\hspace*{-10pt}\frac{\partial \mathcal{L}}{\partial \rho_2} 
	& = \omega_s \bar{\theta} P_s
	- \lambda_1 \bar{\theta} \frac{\frac{P_s \vert h_d \vert ^2}{N}  \left( 1 + \sqrt{\frac{ P_r \vert h_r \vert ^2 \rho_r}{P_s \vert h_d \vert ^2 \rho_2}}  \right)}
		{  \left( 1+ \frac{\left( \eta_2+\rho_2 \right)P_s \vert h_d \vert ^2 + \vert h_r \vert ^2 \rho_r P_r  + 2 \sqrt{P_s \vert h_d \vert ^2 P_r  \vert h_r \vert ^2 \rho_2 \rho_r}}{N}\right)}
=0 \label{eq:lag_rho2}
\\
%\end{align}
%\begin{align}%
\hspace*{-10pt}\frac{\partial \mathcal{L}}{\partial \rho_r} 
	& = \omega_r \bar{\theta} P_r
	- \lambda_1 \bar{\theta} \frac{\frac{P_r \vert h_r \vert ^2}{N} \left( 1 + \sqrt{\frac{P_s \vert h_d \vert ^2  \rho_2}{ P_r \vert h_r \vert ^2\rho_r}}\right)}
		{  \left( 1+ \frac{\left( \eta_2+\rho_2 \right)P_s \vert h_d \vert ^2 + \vert h_r \vert ^2 \rho_r P_r  + 2 \sqrt{P_s \vert h_d \vert ^2 P_r  \vert h_r \vert ^2 \rho_2 \rho_r}}{N}\right)}
=0 \label{eq:lag_rhor}
\end{align}
}
% The spacer can be tweaked to stop underfull vboxes.
%\vspace*{2pt}
% IEEE uses as a separator
\hrulefill
% Restore the current equation number.
%\setcounter{equation}{\value{MYtempeqncnt}}
\end{figure*}

\section{Proof of Algorithm \ref{Algo_N_EE} : Power allocation for the network energy efficiency}
\label{appendix:min_network}

\subsection{General problem setting}
\label{appendix:general}

Denote $X =\left( \eta_1,\rho_1,\eta_2,\rho_2, \rho_r\right)^T$. Consider optimization problem \eqref{eq:min_energy_pbm}. We look for the $X$ that minimizes the objective function $f$, such that
\begin{align*}
f(X) & = \omega_s \left[ \theta \left( \eta_1+\rho_1 \right) P_s + \bar{\theta}\left( \eta_2+\rho_2\right)P_s \right]+ \omega_r\bar{\theta}\rho_r P_r
\end{align*}
given the two rate constraints $c_1(X)= I_1 - R$ and $c_2(X) = I_2 - R$, as defined in \eqref{eq:gauss_const}. Here, we first ignore the power constraints. The Lagrangian can then be formed as
\begin{align*}
\mathcal{L} = f(X) - \lambda_1 c_1(X) - \lambda_2 c_2(X)
\end{align*}
Assuming natural logarithms does not change the optimal solution. The KKT conditions are expressed by equations \eqref{eq:lag_eta1} - \eqref{eq:lag_rhor} at the top of next page and by \eqref{eq:kkt_1}.
\begin{equation} \label{eq:kkt_1}
\begin{split}
\lambda_i c_i(X) = 0 \; ; \quad 
c_i(X) \geq 0 \; ; \quad 
\lambda_i \geq 0 \; ; \quad 
i = 1,2
\end{split}
\end{equation}

%%%%%%%%%%%%%%%%%%%%%%%%%%%%%%%%%%%%%%%%%%%%%%%%%%%%%%%%%%%%%%%%%%%%%%%%%%%%%%%

\subsection{Transmission during the first phase}
\label{appendix:eta1}

Note that $\frac{\partial \mathcal{L}}{\partial \eta_1} $ and $\frac{\partial \mathcal{L}}{\partial \rho_1} $ cannot be equal to zero simultaneously, unless $\vert h_d \vert ^2 = \vert h_s \vert ^2$. Therefore, we have to relax either $\rho_1$ or $\eta_1$. Considering $\vert h_d \vert ^2 < \vert h_s \vert ^2$ for which decode-forward is useful, we get $\frac{\partial \mathcal{L}}{\partial \eta_1}  > \frac{\partial \mathcal{L}}{\partial \rho_1}$. Relaxing $\eta_1$ gives $\frac{\partial \mathcal{L}}{\partial \eta_1}  > 0$ when $\frac{\partial \mathcal{L}}{\partial \rho_1} = 0$, and $\eta_1$ should be minimized. Since the objective function and the first constraint $c_1(X)$ only depend on the sum $\eta_1 + \rho_1$, setting $\eta_1=0$ weakens the second constraint $c_2(X)$ and minimizes $f$. Thus, the source only sends $m_r$ during the first phase.

The following notations will be used for the rest of this appendix and next ones:

\noindent{\small
\begin{align*}
\Gamma_i &= 1 + \frac{\rho_1 P_s \vert h_i \vert ^2}{N} \quad ; \quad G_i  =\frac{2^{R/\bar{\theta}}}{\Gamma_i^{\theta / \bar{\theta}}} \quad \text{where} \; i \in \left\lbrace s,d \right\rbrace  \\
\Gamma_r & = 1+ \frac{\left( \eta_2+\rho_2 \right)P_s \vert h_d \vert ^2 + \vert h_r \vert ^2 \rho_r P_r  + 2 \sqrt{P_s \vert h_d \vert ^2 P_r  \vert h_r \vert ^2 \rho_2 \rho_r}}{N}
\end{align*}}

\subsection{Optimal allocation set for network energy efficiency}

In this case, both $\omega_s$ and $\omega_r$ are equal to 1.
First, from \eqref{eq:lag_rho2} and \eqref{eq:lag_rhor}, we get:

\noindent {\small
\begin{align*}
\lambda_1 & =\frac{\Gamma_r N}{\vert h_d \vert ^2} \left( 1 + \sqrt{\frac{P_r \vert h_r \vert ^2\rho_r}{P_s \vert h_d \vert ^2  \rho_2}}\right)^{-1} 
%\; \text{and} \quad
%\lambda_1  
  =\frac{ \Gamma_r  N}{\vert h_r \vert ^2} \left( 1 + \sqrt{\frac{P_s \vert h_d \vert ^2  \rho_2}{P_r \vert h_r \vert ^2\rho_r}}\right)^{-1} 
\end{align*}
}
\begin{align}
\text{and } \quad \frac{\vert h_d \vert ^2}{\vert h_r \vert ^2} = \sqrt{\frac{P_s \vert h_d \vert ^2 \rho_2}{P_r  \vert h_r \vert ^2 \rho_r}}  \Leftrightarrow
\rho_2 = \frac{P_r \vert h_d \vert^2}{P_s \vert h_r \vert^2} \rho_r . \label{equation_rho2}
\end{align}

As $\lambda_1 > 0$, the first constraint is active and $G_d = \Gamma_r $ gives

\noindent{\small \begin{align}
& \frac{2^{R/\bar{\theta}}}{\left( 1+ \frac{\rho_1 P_s \vert h_d \vert ^2 }{N}\right) ^{\theta/\bar{\theta}}}  =
\nonumber \\
& \quad \quad \quad
 1 + \frac{ \eta_2 P_s \vert h_d \vert ^2 }{N}+ \frac{\left(\sqrt{P_s \vert h_d \vert ^2 \rho_2} + \sqrt{P_r  \vert h_r \vert ^2 \rho_r} \right) ^2 }{N}  \label{equation_rhor} 
\end{align} }
With \eqref{equation_rho2}, we get
\begin{align}
%&
%%\quad \quad \quad
%= 1 + \frac{ \eta_2 P_s \vert h_d \vert ^2 }{N}+ P_r  \vert h_r \vert ^2 \rho_r \left( \sqrt{\frac{P_s \vert h_d \vert ^2 \rho_2}{P_r  \vert h_r \vert ^2 \rho_r}} +1 \right)^2 \nonumber \\
G_d &
%\quad \quad \quad
 = 1 + \frac{ \eta_2 P_s \vert h_d \vert ^2 }{N}+ P_r \rho_r \vert h_r \vert ^2 \left( \frac{\vert h_d \vert ^2}{\vert h_r \vert ^2} +1 \right)^2  \label{equation_rhor_sr}
\end{align}
From \eqref{eq:lag_eta2}, 
%\begin{align*}
$\lambda_2 = \left(1- \frac{1}{1+ \sqrt{\frac{P_r \vert h_r \vert ^2\rho_r}{P_s \vert h_d \vert ^2  \rho_2}} } \right)\left(\frac{ N }{\vert h_d \vert ^2}+  \eta_2 P_s  \right) >0$.
%\end{align*}
Thus, the second constraint is also active and we get
\begin{align}
\frac{ \eta_2 P_s \vert h_d \vert ^2 }{N}
	& = G_s-1 \label{equation_eta2} 
\end{align}

\noindent From \eqref{equation_rho2}, \eqref{equation_rhor_sr} and \eqref{equation_eta2}, we write $\rho_2$, $\rho_r$ and $\eta_2$ as functions of $\rho_1$.
Replacing the Lagrangian multipliers $\lambda_1$ and $\lambda_2$ in \eqref{eq:lag_rho1}, we prove that $\rho_1$ solves $g_1(\rho_1)=0$, where $g_1$ is defined by \eqref{eq:solve_rho_1n} of Algorithm \ref{Algo_N_EE}.
Note that $g_1$ is decreasing in $\rho_1$, increasing in $R$ and that $g_1(0)=\frac{\vert h_s \vert^2 \vert h_r \vert^2 +\vert h_d \vert^4}{\vert h_d \vert^2 \vert h_r \vert^2 +\vert h_d \vert^4} 2^{R / \bar{\theta}} -1 > 0$ for all $R$. If, for all $\rho_1 \in [0, 1/\theta]$, $g_1(\rho_1) > 0$, then the desired source rate is infeasible with the allocation as defined for $A^{(n)}$ and sub-scheme $B^{(n)}$ should be considered instead. 
We can also check that $\rho_r >0$ and $\rho_2 >0$ since $\vert h_s \vert^2 > \vert h_d \vert^2$ by assumption, and $\eta_2 >0$ as long as $R \geq \theta \log_2 \left(1+ \frac{\rho_1 P_s \vert h_s \vert^2}{N} \right)=R^{(n)}$
On the contrary, when $R \leq R^{(n)} $, $\eta_2$ is negative and there is no solution for \eqref{equation_eta2}. Thus, we relax $\eta_2$ and rewrite the optimization. Following the same steps as above, we can show that constraints are still active, and that \eqref{eq:lag_rho1}, \eqref{eq:lag_rho2} and \eqref{eq:lag_rhor} still hold. This leads to Proposition \ref{prop:B_n}. If the power constraints are not satisfied, then the desired source rate is infeasible.

%%%%%%%%%%%%%%%%%%%%%%%%%%%%%%%%%%%%%%%%%%%%%%%%%%%%%%%%%%%%%%%%%%%%%%

\section{Proof of Algorithm \ref{Algo_R_EE} : Power allocation for relay energy efficiency}
\label{appendix:min_relay}

The proof of Algorithm \ref{Algo_R_EE} and Propositions \ref{prop:C_r}, \ref{prop:A_r} and \ref{prop:B_r} follows the same main steps as in Appendix \ref{appendix:min_network}. Here, $\omega_s = 0$ and $\omega_r = 1$. Analysis in Appendices \ref{appendix:general} and \ref{appendix:eta1} still holds. To minimize the energy consumption of the relay, direct transmissions should be performed as long as the direct link is not in outage, or equivalently as long as $R \leq \log_2 \left(1+ \frac{P_s \vert h_d \vert ^2}{N} \right)$ (Proposition \ref{prop:C_r}).

The relay is needed to perform the transmission as soon as the source meets its power constraint and cannot send more data without going in outage. Then,
\begin{align}
\rho_2 = \frac{1-\rho_1 \theta}{\bar{\theta}}-\eta_2 . \label{equation_rho1_r}
\end{align}

First, note that Eq. $\frac{\partial \mathcal{L}}{\partial \rho_r} =0$ \eqref{eq:lag_rhor} is still valid. Thus $\lambda_1 = \Gamma_r \left[ \frac{\vert h_r \vert ^2} {N \ln 2} \left( 1 + \sqrt{\frac{P_s \vert h_d \vert ^2  \rho_2}{ P_r \vert h_r \vert ^2\rho_r}}\right) \right] ^{-1}$.
Second, with \eqref{equation_rho1_r}, the Lagrangian equations now satisfy

\noindent{\small
\begin{align}
\frac{\partial \mathcal{L}}{\partial \eta_2} 
	& = \lambda_1 \frac{\bar{\theta} P_s \vert h_d \vert ^2}{N \Gamma_d \ln 2 }
	- \lambda_1 \bar{\theta} \frac{P_s \vert h_d \vert ^2}
		{N  \ln 2  \Gamma_r}
	+ \lambda_2 \frac{\bar{\theta} P_s \vert h_s \vert ^2}{N \Gamma_s  \ln 2 }
	\nonumber \\
	& \quad %\quad \quad \quad \quad \quad \quad \quad \quad \quad \quad \quad
	 - \lambda_2 \bar{\theta} \frac{P_s \vert h_d \vert ^2 }{N \left( 1+ \frac{\eta_2 P_s \vert h_d \vert ^2 }{N}\right) \ln 2 }
	=0
\label{eq:lag_eta2_relay}\\
\frac{\partial \mathcal{L}}{\partial \rho_2} 
	& = \lambda_1 \frac{\bar{\theta} P_s \vert h_d \vert ^2}{N \Gamma_d \ln 2}
	- \lambda_1 \bar{\theta} \frac{P_s \vert h_d \vert ^2 \left( 1 + \sqrt{\frac{ P_r \vert h_r \vert ^2 \rho_r}{P_s \vert h_d \vert ^2 \rho_2}}  \right)}
		{N  \ln 2  \Gamma_r} 
		\nonumber \\
	& \quad 
+ \lambda_2 \frac{\bar{\theta} P_s \vert h_s \vert ^2}{N \Gamma_s \ln 2 }
		=0
\label{eq:lag_rho2_relay}
\end{align}
}
% \eqref{eq:lag_eta2_relay} and \eqref{eq:lag_rho2_relay} at the bottom of the page,
%  where $\Gamma_i = 1+ \frac{\left(1-(\eta_2 + \rho_2) \bar{\theta}\right) P_s \vert h_i \vert ^2 }{\theta N}$.
Subtracting \eqref{eq:lag_eta2_relay} from \eqref{eq:lag_rho2_relay}, we get

\noindent{\small \begin{align*}
\lambda_2 & =  \lambda_1 \sqrt{\frac{ P_r \vert h_r \vert ^2 \rho_r}{P_s \vert h_d \vert ^2 \rho_2}}\left( 1+ \frac{\eta_2 P_s \vert h_d \vert ^2 }{N}\right) \Gamma_r^{-1}
 \\
	&=\frac{N  }{\vert h_r \vert ^2}  \sqrt{\frac{ P_r \vert h_r \vert ^2 \rho_r}{P_s \vert h_d \vert ^2 \rho_2}} \frac{\left( 1+ \frac{\eta_2 P_s \vert h_d \vert ^2 }{N}\right)} {\left( 1 + \sqrt{\frac{P_s \vert h_d \vert ^2  \rho_2}{ P_r \vert h_r \vert ^2\rho_r}}\right)} >0
\end{align*}
}

Both constraints are active: Eq. \eqref{equation_rhor} $G_d = \Gamma_r \quad$ and Eq. \eqref{equation_eta2} $\frac{ \eta_2 P_s \vert h_d \vert ^2 }{N}  = G_s-1$ still hold. From \eqref{equation_eta2}, $\eta_2$ is expressed as a function of $\rho_1$. From \eqref{equation_rhor}, \eqref{equation_rho1_r} and the expression of $\eta_2$, $\rho_r$ can be also written as a function of $\rho_1$.

Finally, plugging the expressions of $\lambda_1$ and $\lambda_2$ in \eqref{equation_rhor}, \eqref{equation_eta2} into \eqref{eq:lag_rho2_relay}, we prove that $\rho_1$ should satisfy $g_2(\rho_1)=0$, as defined by \eqref{eq:solve_rho_1r} of Algorithm \ref{Algo_R_EE}, with

\noindent{\small \begin{align*}
g_3(\rho_1) &= \sqrt{\frac{P_r \vert h_r \vert ^2 \rho_r}{P_s \vert h_d \vert ^2 \rho_2}}\\
&= \frac{2^{R/ \bar{\theta}}}{\left( 1+\frac{\rho_1 P_s \vert h_d \vert ^2}{ N}\right)^{\theta / \bar{\theta}}}- \frac{2^{R/ \bar{\theta}}}{\left( 1+ \frac{\rho_1 P_s \vert h_s \vert ^2 }{ N} \right)^{\theta / \bar{\theta}}} 
\end{align*}
}

Now, we focus on the existence of this power allocation set. First, to ensure $\eta_2 >0$, $R$ must be greater than $\theta \log_2 \left( 1+\frac{\rho_1 Ps \vert h_s \vert ^2}{N}\right)=R^{(r)}_2$. Moreover, $\rho_r$ is positive as long as
%%\begin{align*}
$R > \theta \log_2 \left( 1+ \frac{\rho_1 P_s \vert h_d \vert ^2}{N}\right) + \bar{\theta} \log_2 \left( 1+ \frac{\left( \eta_2+\rho_2 \right)P_s \vert h_d \vert ^2}{N}\right)
%\\ &
= \log_2 \left(1+ \frac{P_s \vert h_d \vert ^2}{N} \right) =R^{(r)}_1 $.
%\end{align*}
Finally, the existence of $\rho_1$, and consequently of $\rho_2$, follows the same steps as in Appendix \ref{appendix:min_network}. This concludes the proof of Proposition \ref{prop:A_r}.

When $R<\theta \log_2 \left( 1+\frac{\rho_1 Ps \vert h_s \vert ^2}{N}\right)$, $\eta_2$ is negative and there is no solution for \eqref{equation_eta2}. In this case, we relax $\eta_2$ and rewrite the optimization problem. As above, we can show that rate constraints are still active, and that \eqref{eq:lag_rhor} and \eqref{eq:lag_rho2_relay} still hold.
This concludes the proof of Proposition \ref{prop:B_r} and Algorithm \ref{Algo_R_EE}.

\section{Proof of Algorithm \ref{Algo_S_EE} : Power allocation for source energy efficiency}
\label{appendix:min_source}

The proof of Algorithm \ref{Algo_S_EE} and Propositions \ref{prop:C_s}, \ref{prop:B_s} and \ref{prop:A_s} follows the same steps as in Appendix \ref{appendix:min_network}. Here, $\omega_s = 1$ and $\omega_r = 0$. Analysis of Lagrangian in \ref{appendix:general} and \ref{appendix:eta1} still holds.
However, the constraint on the relay consumption is relaxed here, such that it can consume up to its maximal power. %Two cases should thus be considered: 1) the relay has to use all its available power to transmit data, 2) the relay is not limiting.

First, let's consider that the source rate is high and that all the available relay power is required ($\rho_r = \frac{1}{\bar{\theta}}$). This means that the first rate constraint, which depends on $\rho_r$ will necessarily be tight. From Eq. $\frac{\partial \mathcal{L}}{\partial \rho_1} =0$ \eqref{eq:lag_rho1} and Eq. $\frac{\partial \mathcal{L}}{\partial \rho_2} =0$ \eqref{eq:lag_rho2}, we can prove that both Lagrangian multipliers are positive, meaning that both constraints are active, such that

\noindent{\small
\begin{align*}
\lambda_1 & = \frac{ \Gamma_r} {\frac{\vert h_d \vert ^2}{ N} \left( 1 + \sqrt{\frac{P_r \vert h_r \vert ^2\rho_r}{P_s \vert h_d \vert ^2  \rho_2}}\right)}
% \\
%& 
= G_d \frac{ N}{\vert h_d \vert ^2  \left( 1 + \sqrt{\frac{P_r \vert h_r \vert ^2\rho_r}{P_s \vert h_d \vert ^2  \rho_2}}\right)} \\
%\end{align*}
%%
%\begin{align*}
\lambda_2 & = \left(1- \frac{\lambda_1  \vert h_d \vert ^2}{N \ln 2 \Gamma_d} \right) \frac{N \ln 2 \Gamma_s}{\vert h_s \vert ^2}
 \\
&
= \left(1- \frac{G_d}{\left( 1 + \sqrt{\frac{ P_r \vert h_r \vert ^2\rho_r}{P_s \vert h_d \vert ^2  \rho_2}}\right)} \right)
% \\
%&\qquad \qquad \qquad \qquad\qquad \qquad
%\times
\frac{N \ln 2 \Gamma_s}{\vert h_s \vert ^2}
\end{align*}
}
Plugging in these expressions of $\lambda_1$ and $\lambda_2$, as well as Eq. $G_d = \Gamma_r \quad$ \eqref{equation_rhor} and Eq. $\frac{ \eta_2 P_s \vert h_d \vert ^2 }{N}  = G_s-1$ \eqref{equation_eta2} into \eqref{eq:lag_rho2}, we prove that $\rho_1$ should satisfy $g_2(\rho_1)=0$, as defined by \eqref{eq:solve_rho_1s} of Algorithm \ref{Algo_S_EE}. The existence of this power allocation set follows the same steps as for Appendices \ref{appendix:min_network} and \ref{appendix:min_relay}. This leads to sub-scheme $A^{(s)}$ and Proposition \ref{prop:A_s}. Note that the power allocation is similar to sub-scheme $A^{(r)}$.
We now have to insure the non-negativity of this solution, in particular $\eta_2 \geq 0$ and $\rho_2 \geq 0$.

On the one hand, we consider the non-negativity constraint on $\eta_2$. $\eta_2$ is positive following sub-scheme $A^{(s)}$ as long as $R \geq R^{(s)}_{2,B}$ with $R^{(s)}_{2,B} = \theta \log_2 \left(1+ \frac{\rho_1^\star P_s \vert h_s \vert^2}{N} \right)$. If $R \leq R^{(s)}_{2,B}$, $\eta_2$ is relaxed and set to 0. Next, the optimization follows the same steps as for $B^{(n)}$ or $B^{(r)}$, but with $\rho_r = \frac{1}{\bar{\theta}}$. This leads to sub-scheme $B^{(s)}$ and Proposition \ref{prop:B_s}.
Note that, following $B^{(s)}$, $\rho_2$ is positive as long as $R \geq R^{(s)}_{1,B}$, with $R^{(s)}_{1,B} = R_B$ as defined in Algorithm \ref{Algo_S_EE}. This means that the relay requires all its power to forward the source data such that the first rate constraint $c_1$ is active. However, if the rate decreases under the threshold $R^{(s)}_{1,B}$, the relay may not need all its available power and the first rate constraint becomes inactive. $\rho_2$ is relaxed and set to 0.
Finally, even if the relay consumption is not part of the optimization problem here, we can nevertheless reduce it to the minimum required, such that $R = R^{(s)}_1$ (however, $c_1$ is still inactive in the sense of Lagrangian optimization). Thus, we deduce the expression of $\rho_r < 1/\bar{\theta}$.
The optimization now leads to sub-scheme $D^{(s)}$ and Proposition \ref{prop:D_s}.

On the other hand and following sub-scheme $A^{(s)}$ again, we consider the non-negativity constraint on $\rho_2$ (with $\eta_2>0$). $\rho_2$ is positive following sub-scheme $A^{(s)}$ as long as $R \geq R^{(s)}_{2,C}$ with $R^{(s)}_{2,C} = \bar{\theta} \log_2 \left(\frac{P_r \vert h_r \vert^2}{\bar{\theta}N} \right) - \bar{\theta} \log_2 \left(\frac{1}{\left( 1+\frac{\rho_1^\star P_s \vert h_d \vert ^2}{ N}\right)^{\theta / \bar{\theta}}} -\frac{1}{\left( 1+\frac{\rho_1^\star P_s \vert h_s \vert ^2}{ N}\right)^{\theta / \bar{\theta}}} \right)$.

\noindent If $R \leq R^{(s)}_{2,C}$, $\rho_2$ is relaxed and set to 0. This also means that the relay is able to forward the source data without requiring the beamforming gain between the source and the relay. Therefore, the first rate constraint $c_1$ becomes inactive and $\lambda_1 = 0$. From Eq. $\frac{\partial \mathcal{L}}{\partial \rho_1} =0$ \eqref{eq:lag_rho1} and Eq. $\frac{\partial \mathcal{L}}{\partial \eta_2} =0$ \eqref{eq:lag_eta2}, we get
$\frac{\vert h_s \vert ^2}{1+ \frac{\rho_1 P_s \vert h_s \vert^2}{N}} = \frac{\vert h_d \vert ^2}{1+ \frac{\eta_2 P_s \vert h_d \vert^2}{N}}$, such that $\eta_2  = \rho_1 + \frac{N}{P_s}\left( \frac{1}{\vert h_s \vert^2}- \frac{1}{\vert h_d \vert^2} \right)$. 
As the second constraint $c_2$ is active, Eq. $\frac{ \eta_2 P_s \vert h_d \vert ^2 }{N}  = G_s-1$ \eqref{equation_eta2} is still valid and we get $\eta_2 = \frac{N}{P_s \vert h_d \vert ^2} \left( \frac{2^{R/ \bar{\theta}}}{\left( 1+\frac{\rho_1 P_s \vert h_s \vert ^2}{N}\right)^{\theta / \bar{\theta}}}- 1\right)$. By equalizing both expressions of $\eta_2$, we deduce $\rho_1$. Next, even if the relay consumption is not considered here, we reduce it to its minimum required value. The optimization leads to sub-scheme $C^{(s)}$ and Proposition \ref{prop:C_s}.
Now, following $C^{(s)}$, $\eta_2$ is positive as long as $R \geq R^{(s)}_{1,C}$, with $R^{(s)}_{1,C} = R_C$ as defined in Algorithm \ref{Algo_S_EE}. If the rate decreases under the threshold $R^{(s)}_{1,C}$, $\eta_2$ is relaxed and set to 0, which leads to $D^{(s)}$ as previously defined.

So far, depending on non-negativity constraints, we defined two possible sequences of sub-schemes: $(A^{(s)},B^{(s)},D^{(s)})$ and $(A^{(s)},C^{(s)},D^{(s)})$. We now define how each of them should be used. 
First, $A^{(s)}$ is applied when $R \geq \max \left \lbrace R^{(s)}_{2,B} , R^{(s)}_{2,C} \right \rbrace = R^{(s)}_2$. If $R^{(s)}_{2,C} \leq R^{(s)}_{2,B}$, the non-negativity constraint on $\eta_2$ is stronger than the constraint on $\rho_2$ and the sequence $(A^{(s)},B^{(s)})$ is applied. Otherwise, $(A^{(s)},C^{(s)})$ is applied.
Second, $D^{(s)}$ is such that $\eta_2=0$ and $\rho_2=0$. This sub-scheme is applied as long as $R \leq \min \left \lbrace  R^{(s)}_{1,B}, R^{(s)}_{1,C}\right \rbrace = R^{(s)}_1$. Thus, if $R^{(s)}_{1,B} \leq R^{(s)}_{1,C}$, the sequence $(B^{(s)}, D^{(s)})$ is applied, otherwise $(C^{(s)}, D^{(s)})$ is applied.

Finally, we show that upper and lower rate bounds for applying either $B^{(s)}$ or $C^{(s)}$ are consistent. Considering the source consumption as a function of the source rate, sub-schemes are continuous. Moreover, $\eta_2$ and $\rho_2$ are increasing functions of the source rate. Thus, if $R^{(s)}_{1,B} \leq R^{(s)}_{1,C}$ and sequence $(D^{(s)}, B^{(s)})$ is applied for low source rates, we get $\rho_2 > 0$ for any $R > R^{(s)}_{1,B}$. In this case, we necessarily have $R > R^{(s)}_{2,C}$ which implies $R^{(s)}_{2,C} \leq R^{(s)}_{2,B}$.  Therefore, only sequence $(A^{(s)},B^{(s)},D^{(s)})$ can occur. Similarly, we can show that only $(A^{(s)},C^{(s)},D^{(s)})$ can occur if $R^{(s)}_{1,B} \geq R^{(s)}_{1,C}$.
This completes the proof of Algorithm \ref{Algo_S_EE}.

%
%If $R \geq R^{(s)}_2$ but $R \leq R^{(s)}_1$ (Proposition \ref{prop:C_s}), the first constraint $c_1$ is inactive and $\lambda_1 = 0$. Thus, $\rho_2$ is relaxed and set to 0. From \eqref{eq:lag_rho1} and \eqref{eq:lag_eta2}, we get
%$\frac{\vert h_s \vert ^2}{1+ \frac{\rho_1 P_s \vert h_s \vert^2}{N}} = \frac{\vert h_d \vert ^2}{1+ \frac{\eta_2 P_s \vert h_d \vert^2}{N}}$. 
%As the second constraint $c_2$ is active, \eqref{equation_eta2} is still valid and we can deduce $\rho_1$ and $\eta_2$.
%Finally, even if the relay consumption is not considered here, we can nevertheless reduce it to the minimum, such that $R = R^{(s)}_1$ (however, $c_1$ is still inactive in the sense of Lagrangian optimization). Thus, we deduce the expression of $\rho_r < 1/\bar{\theta}$.
%
%Finally, if $R < R^{(s)}_2$ but $R < R^{(s)}_1$ (Proposition \ref{prop:D_s}), both $\rho_2$ and $\eta_2$ should be relaxed and set to 0.
%
%$ \frac{N}{P_s \vert h_d \vert ^2} \left(2^R \left( \frac{\vert h_d \vert ^2}{\vert h_s \vert ^2}\right)^\theta-1 \right) = \frac{N}{P_s \vert h_d \vert ^2} \left(2^R \left( \frac{\vert h_d \vert ^2}{\vert h_s \vert ^2}\right)^\theta-1 \right) $

%
%This concludes the proof of Theorem \ref{th:s}.

\section{Maximal achievable source rate}
\label{appendix:Maximal_achievable_source_rate}

\subsection{R-EE and S-EE achieve higher rates than N-EE}
\label{appendix:energy_same_max_rate_N_EE}

A source rate R is achievable for N-EE (resp. R-EE) as long as there exists a $\rho_1^\star$ satisfying $g_1(\rho_1^\star)=0$ (resp. $g_2(\rho_1^\star)=0$). Recall that $g_1(0)>0$, $g_2(0)>0$ and that both functions are decreasing (see Appendix \ref{appendix:min_network}). Since $\forall x, \; g_2(x)<g_1(x)$, there exists a range of $\rho_1$ for which $g_1(\rho_1)>0$ but $g_2(\rho_1)\leq 0$. Therefore, for this range, we can find a solution for $g_2(\rho_1^\star)=0$ but not for $g_1(\rho_1^\star)=0$. The corresponding range of source rates is thus achieved with R-EE, but not with N-EE. The proof is similar for S-EE.

\subsection{Derivation of $R_{\max}$ with sub-schemes $D^{(s)}$ and $C^{(s)}$} \label{appendix:energy_same_max_rate_2}

If sub-scheme $D^{(s)}$ is applied when S-EE goes in outage, the source power constraint is met, such that $(\rho_1^\circ+\eta_1^\circ) \theta P_s + (\rho_2^\circ+\eta_2^\circ) \bar{\theta} P_s = P_s$. This leads to $\rho_1^\circ = \frac{1}{\theta}$ and $R_{\max} = \theta \log_2 \left(1+ \frac{P_s \vert h_s \vert^2}{\theta N} \right)$.

Similarly, if sub-scheme $C^{(s)}$ is applied when S-EE goes in outage, $(\rho_1^\ddagger+\eta_1^\ddagger) \theta P_s + (\rho_2^\ddagger+\eta_2^\ddagger) \bar{\theta} P_s = P_s$.
Thus, $\theta \rho_1^\ddagger + \bar{\theta} \left( \rho_1^\ddagger + \frac{N}{P_s}\left( \frac{1}{\vert h_s \vert^2}- \frac{1}{\vert h_d \vert^2} \right)\right) = 1$
and $\rho_1^\ddagger = 1+\frac{\bar{\theta} N}{P_s} \left(\frac{1}{\vert h_d \vert 2} - \frac{1}{\vert h_s \vert 2} \right) =\frac{N}{P_s \vert h_s \vert ^2} \left( 2^R \left( \frac{\vert h_s \vert ^2}{\vert h_d \vert ^2}\right)^{\bar{\theta}} -1\right)$, from which we deduce $R_{\max}$ as in Section \ref{prop:scheme_comp}.

\subsection{S-EE and R-EE achieve the same maximum rate $R_{\max}$}
\label{appendix:energy_same_max_rate_3}

We want to prove that the power allocation of S-EE for $R=R_{\max}$ corresponds to the allocation of R-EE at this specific rate such that both achieve the same maximum rate $R_{\max}$.
Also note that as long as $\vert h_s \vert^2 > \vert h_d \vert^2$, the maximum rate achieved by R-EE using $C^{(r)}$ (direct transmission) is strictly lower to the maximum rate achieved by R-EE using $B^{(r)}$ or $A^{(r)}$. Thus, when R-EE declares outage, either sub-scheme $B^{(r)}$ or $A^{(r)}$ is applied.
Now, we successively consider the four possible cases for S-EE.

First, assume that $A^{(s)}$ is applied when S-EE goes in outage at $R=R_{\max}$. This means that $R \geq \theta \log_2 \left(1+ \frac{\rho_1 P_s \vert h_s \vert^2}{N} \right)$ where $\rho_1$ solves $g_2(\rho_1)=0$. Therefore, R-EE uses $A^{(r)}$ at this rate and $\rho_1^{(r)}=\rho_1^{(s)}$. 
Now, note that $\eta_2^{(r)}=\eta_2^{(s)}$, since both solve the same expression. For R-EE, $\rho_2^{(r)}= \frac{1-\rho_1^{(r)}\theta}{\bar{\theta}}$ and S-EE declares outage when the source power constraint is met. Thus, $\rho_2^{(r)}= \rho_2^{(s)}$ at $R=R_{\max}$. Noting that $\rho_2$ and $\rho_r$ satisfy the same equation in both R-EE and S-EE
%$1+ \frac{\left( \eta_2+\rho_2 \right)P_s \vert h_d \vert ^2 + \vert h_r \vert ^2 \rho_r P_r  + 2 \sqrt{P_s \vert h_d \vert ^2 P_r  \vert h_r \vert ^2 \rho_2 \rho_r}}{N} = \frac{2^{R/\bar{\theta}}}{\left( 1 + \frac{\rho_1 P_s \vert h_d \vert ^2}{N} \right)^{\theta / \bar{\theta}}}$
($\Gamma_r = G_d$), we get $\rho_r^{(r)} = \rho_r^{(s)} = 1/ \bar{\theta}$, which means that the relay constraint is also achieved and R-EE goes in outage for this same source rate. 
%
% For S-EE, $\rho_r^{(s)}= \frac{1}{\bar{\theta}}$ and R-EE declares outage when the relay power constraint is met. Thus, at this specific rate, $\rho_r^{(r)}=\rho_r^{(s)}$. Similarly, f
%
%If $\rho_2^{(r)}=\rho_2^{(s)}$ and $\rho_r^{(r)}=\rho_r^{(s)}$, we can easily check that $\eta_2^{(r)}=\eta_2^{(s)}$ and that $\rho_1^{(r)}=\rho_1^{(s)}$ since $g_2 =g_3$ in this case.
%The maximum achievable rate of S-EE is achieved when the source power constraint is met with equality, which leads to $\rho_2^{(s)} = \frac{\left(1-\rho_1^{(s)} \theta \right)}{\bar{\theta}} - \eta_2^{(s)} = \rho_2^{(r)}$. 

Second, assume that $B^{(s)}$ is used when S-EE goes in outage. This means that $R \leq \theta \log_2 \left(1+ \frac{\rho_1 P_s \vert h_s \vert^2}{N} \right)$ where $\rho_1$ solves $g_2(\rho_1)=0$ and that R-EE uses $B^{(r)}$ when outage occurs. Similarly to the above analysis, we can show that the power allocation of S-EE for $R=R_{\max}$ corresponds to the allocation of R-EE at this specific rate. 

Third, we assume that $C^{(s)}$ is used when S-EE goes in outage. In this case, we have, by concavity,

\noindent{\small
\begin{align*}
R_{\max} & = \log_2 \left(\theta \left(1 + \frac{P_s \vert h_s \vert ^2}{N \theta}  \right) + \bar{\theta}  \frac{\vert h_s \vert ^2}{\vert h_d \vert ^2} \right) - \bar{\theta} \log_2 \left( \frac{\vert h_s \vert ^2}{\vert h_d \vert ^2} \right) \\
			& \geq \theta \log_2 \left(1 + \frac{P_s \vert h_s \vert ^2}{N \theta}  \right) +  \bar{\theta} \log_2 \left( \frac{\vert h_s \vert ^2}{\vert h_d \vert ^2} \right) - \bar{\theta} \log_2 \left( \frac{\vert h_s \vert ^2}{\vert h_d \vert ^2} \right)\\
			& = \theta \log_2 \left(1 + \frac{P_s \vert h_s \vert ^2}{N \theta}  \right) 
\end{align*} 
}
Thus, for all $\rho_1 \leq \frac{1}{\theta}$ solving $g_2(\rho_1)=0$, we get $R_{\max} \geq \theta \log_2 \left(1 + \frac{ \rho_1 P_s \vert h_s \vert ^2}{N}  \right)$ such that at $R =R_{\max}$, R-EE uses $A^{(r)}$. Note that S-EE declares outage using $C^{(s)}$ if the source rate is met, such that $\eta_2^{(s)} = \frac{\left(1-\rho_1^{(s)} \theta \right)}{\bar{\theta}}$. Thus, we can show that the allocation of S-EE is also a solution for $A^{(r)}$ at this specific source rate.
Similar analysis can be done for $D^{(s)}$.

Therefore, for all cases, the power allocation of S-EE for $R=R_{\max}$ corresponds to the allocation of R-EE at this specific rate and both S-EE and R-EE achieve the same maximum rate $R_{\max}$.

\section{Analysis of scheme G-EE: Continuity and Differentiability}

%\subsection{Continuity and Differentiability}
\label{appendix:diff}

First, we show the continuity of both schemes N-EE and R-EE at $R = R^{(n)}_{\max}$. Note that $\eta_1$ and $\eta_2$ solve the same equations for both sub-schemes $A^{(n)}$ and $A^{(r)}$. Then, since the first rate constraint is active, we get $g_3(\rho_1^\star) =  \sqrt{\frac{P_r \vert h_r \vert ^2 \rho_r^\star}{P_s \vert h_d \vert ^2 \rho_2^\star}}$. In $A^{(n)}$, $g_3(\rho_1^\star) = \frac{\vert h_r \vert ^2}{\vert h_d \vert ^2}$, such that $\rho_1^\star$ and $\rho_r^\star$ of $A^{(n)}$ are also optimal for $A^{(r)}$. Finally, at $R = R^{(n)}_{\max}$, the source power constraint is met, such that $\rho_2^\star = \frac{\left(1-\rho_1^\star \theta \right)}{\bar{\theta}} - \eta_2^\star $ for both sub-schemes. Therefore, the optimal power allocation set for $A^{(n)}$ is also optimal for $A^{(r)}$ at this particular rate and schemes N-EE and R-EE are continuous.

Second, note that parameters $\rho_1$, $\eta_2$, $\rho_2$ and $\rho_r$ are not differentiable themselves at $R = R^{(n)}_{\max}$, but that the energy of G-EE (the weighted sum of $\rho_1$, $\eta_2$, $\rho_2$ and $\rho_r$) is differentiable. To prove it, let's consider the left and right derivatives of G-EE. The left derivative (resp. the right derivative) corresponds to the derivative of the total energy consumed by N-EE (resp. R-EE) at $R = R^{(n)}_{\max}$. If both semi-derivatives are equal, then G-EE is differentiable at this source rate. Let's write both as functions of $\frac{\partial \rho_1}{\partial R}$, $\frac{\partial G_s}{\partial R}$ and $\frac{\partial G_d}{\partial R}$. Also note that $\frac{\partial G_i}{\partial R} = \frac{\ln 2}{\bar{\theta}} G_i - \frac{\theta P_s \vert h_s \vert ^2}{\bar{\theta} N_0} \frac{	G_i}{\Gamma_i} \frac{\partial \rho_1}{\partial R}$.

Let's consider the left derivative (scheme N-EE). In this case, we have:

\noindent{\small
\begin{align*}
\frac{\partial E_{\text{N-EE}}}{\partial R} &= \theta	P_s \frac{\partial \rho_1^{(n)}}{\partial R} + \bar{\theta} P_s \left( \frac{\partial \eta_2^{(n)}}{\partial R}+ \frac{\partial \rho_2^{(n)}}{\partial R}\right) + \bar{\theta} P_r \frac{\partial \rho_r^{(n)}}{\partial R}
\\ &
= \theta	P_s \frac{\partial \rho_1^{(n)}}{\partial R} + \left(\frac{N \bar{\theta}}{\vert h_d \vert ^2} - \frac{N \bar{\theta}}{\vert h_d \vert ^2 + \vert h_r \vert ^2} \right)  \frac{\partial G_s^{(n)}}{\partial R}
\\ &
+ \frac{N \bar{\theta}}{\vert h_d \vert ^2 + \vert h_r \vert ^2} \frac{\partial G_d^{(n)}}{\partial R}
\end{align*}
}

Since the source power constraint is met with equality, the right derivative (scheme R-EE) is such that

\noindent{\small
\begin{align*}
\frac{\partial E_{\text{R-EE}}}{\partial R} &=  \bar{\theta} P_r \frac{\partial \rho_r^{(r)}}{\partial R} 
=
\frac{ \bar{\theta} N}{\vert h_r \vert ^2} \left( 
\frac{P_s \vert h_d \vert ^2}{N} \frac{\partial \rho_2^{(r)}}{\partial R}
+ \frac{\partial G_d^{(r)}}{\partial R} - \frac{\partial G_d^{(r)}}{\partial R} \right. \\
& \left.- \sqrt{\frac{P_s \vert h_d \vert ^2}{N}} \left(
 \sqrt{\frac{\left( G_d-G_s\right)}{\rho_2} } \frac{\partial \rho_2^{(r)}}{\partial R}
 \right. \right. \\ & \left. \left.
+ \sqrt{\frac{\rho_2} {\left( G_d-G_s\right)} }\left( \frac{\partial G_d^{(r)}}{\partial R} - \frac{\partial G_d^{(r)}}{\partial R} \right)
\right) \right)
\end{align*}
}

Since $\frac{\partial \rho_2^{(r)}}{\partial R} = - \frac{\theta}{\bar{\theta}} \frac{\partial \rho_1^{(r)}}{\partial R} - \frac{N}{P_s \vert h_d\vert^2} \frac{\partial G_s^{(r)}}{\partial R}$, then

\noindent{\small
\begin{align*}
\frac{\partial E_{\text{R-EE}}}{\partial R} &= 
\frac{ \bar{\theta} N}{\vert h_r \vert ^2} \left( \frac{P_s \vert h_d \vert ^2}{N} - \sqrt{\frac{P_s \vert h_d \vert ^2}{N}\frac{\left( G_d-G_s\right)}{\rho_2} }\right) \frac{\partial \rho_1^{(r)}}{\partial R}
\\&
+\frac{ \bar{\theta} N}{\vert h_r \vert ^2} \left( 1 - \sqrt{\frac{P_s \vert h_d \vert ^2}{N}\frac{\rho_2} {\left( G_d-G_s\right)} }\right) \frac{\partial G_d^{(r)}}{\partial R} \\
& -\frac{ \bar{\theta} N} {\vert h_r \vert ^2} \left(
\frac{N}{P_s \vert h_d \vert ^2} \left( \frac{P_s \vert h_d \vert ^2}{N} - \sqrt{\frac{P_s \vert h_d \vert ^2}{N}\frac{\left( G_d-G_s\right)}{\rho_2} }\right)
\right. \\& \left.
+ \left( 1 - \sqrt{\frac{P_s \vert h_d \vert ^2}{N}\frac{\rho_2} {\left( G_d-G_s\right)} }\right) 
\right) \frac{\partial G_s^{(r)}}{\partial R} 
\end{align*}
}

Now, recall that both N-EE and R-EE are continuous at this particular rate. Thus, $\rho_2 = \frac{P_r \vert h_d \vert^2}{P_s \vert h_r \vert^2} \rho_r = \frac{N \vert h_d \vert ^2 \left(G_d - G_s \right)}{P_s \left( \vert h_r \vert^2+\vert h_d \vert^2 \right)^2 } $ and the right derivative of G-EE is also such that

\noindent{\small
\begin{align*}
\frac{\partial E_{\text{R-EE}}}{\partial R} = &
 \theta	P_s \frac{\partial \rho_1^{(r)}}{\partial R} + \left(\frac{N \bar{\theta}}{\vert h_d \vert ^2} - \frac{N \bar{\theta}}{\vert h_d \vert ^2 + \vert h_r \vert ^2} \right)  \frac{\partial G_s^{(r)}}{\partial R}
 \\ & + \frac{N \bar{\theta}}{\vert h_d \vert ^2 + \vert h_r \vert ^2} \frac{\partial G_d^{(r)}}{\partial R}
\end{align*}
}
which is similar to the left derivative $\frac{\partial E_{\text{N-EE}}}{\partial R}$. Now, define the constant $C$ (that may depend on channel gains or on the source rate) such that $\frac{\partial \rho_1^{(r)}}{\partial R} = \frac{\partial \rho_1^{(n)}}{\partial R} + C$. We get $\frac{\partial G_i^{(r)}}{\partial R} = \frac{\partial G_i^{(n)}}{\partial R} - C \frac{\theta P_s \vert h_i \vert ^2}{\bar{\theta} N} \frac{	G_i}{\Gamma_i}$ and 

\noindent{\small
\begin{align*}
\frac{\partial E_{\text{R-EE}}}{\partial R} & = \frac{\partial E_{\text{N-EE}}}{\partial R} - C \theta P_s \left(
\left( 1- \frac{\vert h_d \vert ^2}{\vert h_d \vert ^2+\vert h_r \vert ^2}\right) \frac{\vert h_s \vert ^2}{\vert h_d \vert ^2}
		\frac{G_s}{\Gamma_s}
		\right.\\ & \left.
		+ \frac{\vert h_d \vert ^2}{\vert h_d \vert ^2+\vert h_r \vert ^2}
		\frac{G_d}{\Gamma_d}-1
\right) \\
& =  \frac{\partial E_{\text{N-EE}}}{\partial R} - C \theta P_s \left(1+\frac{\vert h_r\vert ^2}{\vert h_d\vert ^2} \right) g_1(\rho_1)
\end{align*}
}
Since $g_1(\rho_1) =0$, then $\frac{\partial E_{\text{N-EE}}}{\partial R} = \frac{\partial E_{\text{R-EE}}}{\partial R}$ at $R = R^{(n)}_{\max}$, which completes the proof.

%\subsection{Comparison with maximum rate scheme}
%\label{rate_max}

\bibliographystyle{IEEEtranN}
{\footnotesize 
\bibliography{references}
}

\begin{IEEEbiography}[{\includegraphics[width=1in,height=1.25in,clip,keepaspectratio]{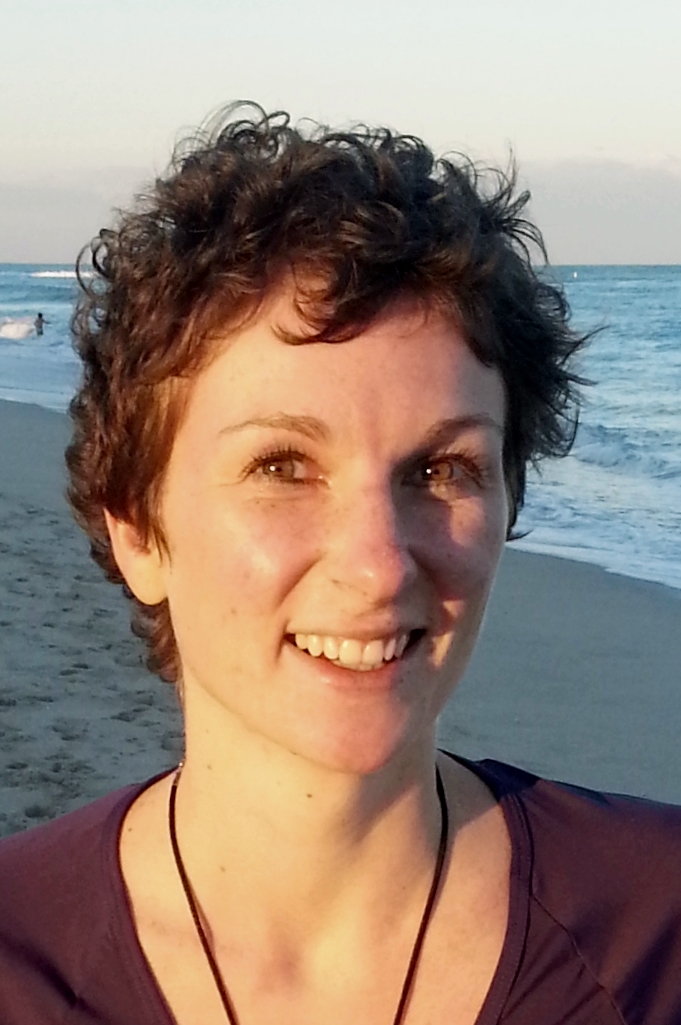}}]{Fanny Parzysz}
received the M.Sc. degree in digital communications and cellular networks from Telecom ParisTech (ENST), Paris, France, in 2009. She is currently pursuing the Ph.D. degree at the LACIME Laboratory, \'{E}cole de Technologie Sup\'{e}rieure, Montreal, Canada. Her Ph.D. is part of the NSERC-Ultra Electronics Industrial Chair in Wireless Emergency and Tactical Communication. Her research interest covers information theory, coding and resource allocation for relay networks, with a particular focus on energy efficiency.
\end{IEEEbiography}
\vfill
\newpage
\begin{biography}[{\includegraphics[width=1in,height=1.25in,clip,keepaspectratio]{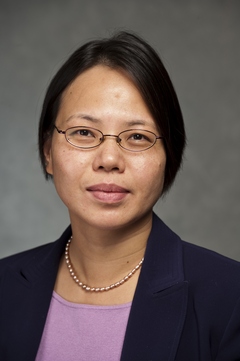}}]{Mai Vu}
received a PhD degree in Electrical Engineering from Stanford University after having an MSE degree in Electrical Engineering from the University of Melbourne and a bachelor degree in Computer Systems Engineering from RMIT, Australia. Between 2006-2008, she worked as a lecturer and researcher at the School of Engineering and Applied Sciences, Harvard University. During 2009-2012, she was an assistant professor in Electrical and Computer Engineering at McGill University. Since January 2013, she has been an associate professor in the department of Electrical and Computer Engineering at Tufts University.

Dr. Vu conducts research in the general areas of wireless communications, signal processing for communications, network communications and information theory. Examples include cooperative and cognitive communications, relay networks, MIMO systems. Dr. Vu has served on the technical program committee of numerous IEEE conferences and is
currently an editor for the IEEE Transactions on Wireless Communications. 
\end{biography}

\begin{IEEEbiography}[{\includegraphics[width=1in,height=1.25in,clip,keepaspectratio]{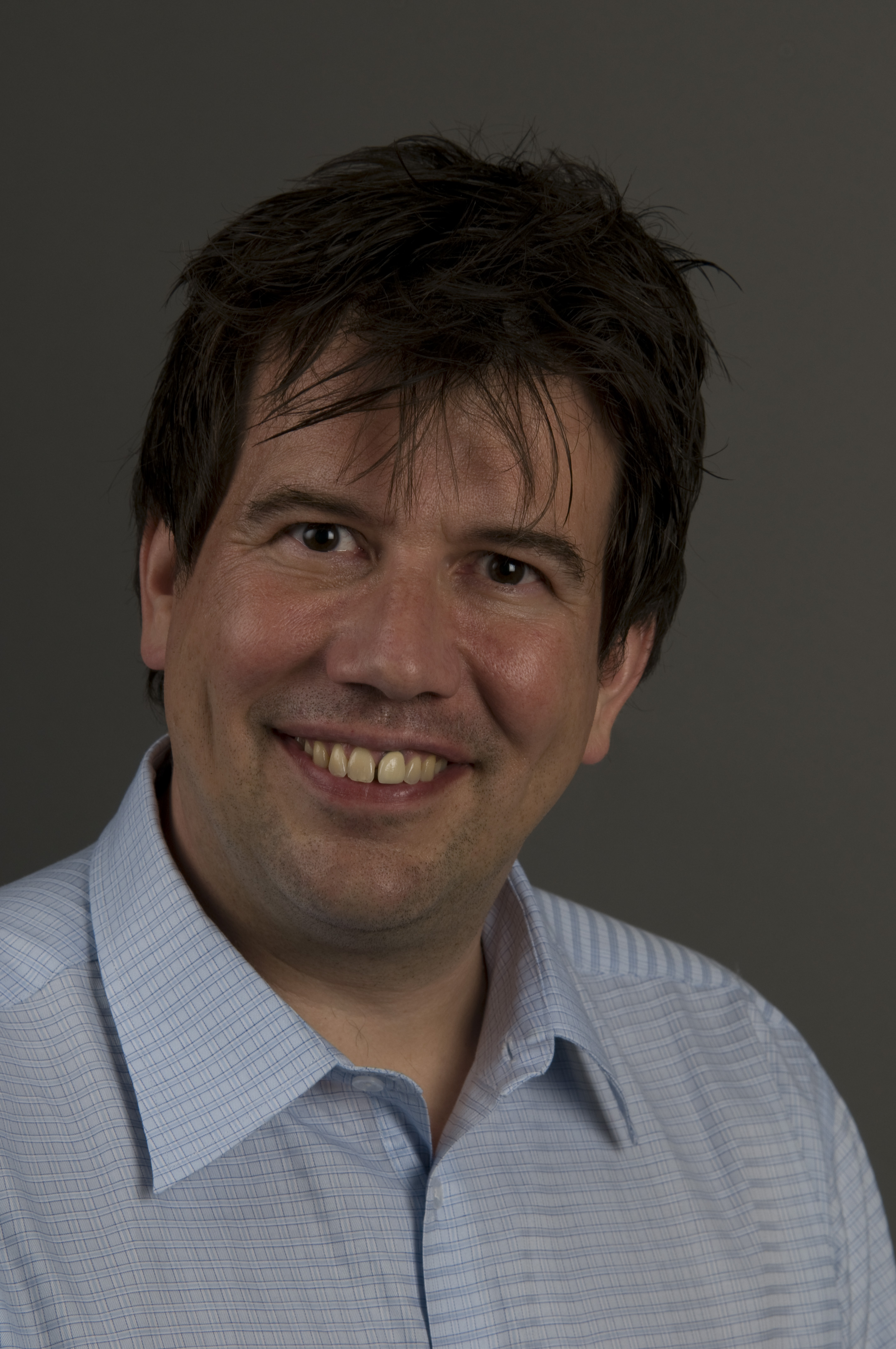}}]{Fran\c{c}ois Gagnon}
received the B.Eng. and Ph.D. degrees in electrical engineering from  \'{E}cole Polytechnique de Montr\'{e}al, Montreal, Quebec, Canada. Since 1991, he has been a Professor with the Department of Electrical Engineering,  \'{E}cole de Technologie Sup\'{e}rieure, Montreal, Quebec, Canada. He chaired the department from 1999 to 2001, and is now the holder of the NSERC Ultra Electronics Chair, Wireless Emergency and Tactical Communication, at the same university. His research interest covers wireless high-speed communications, modulation, coding, high-speed DSP implementations, and military point-to-point communications. He has been very involved in the creation of the new generation of high-capacity line of-sight military radios offered by the Canadian Marconi Corporation, which is now Ultra Electronics Tactical Communication Systems. The company has received, for this product, a “Coin of Excellence” from the U.S. Army for performance and reliability. Prof. Gagnon was awarded the 2008 NSERC Synergy Award for the fruitful and long lasting collaboration with Ultra Electronics TCS.
\end{IEEEbiography}
\vfill

\end{document}